\documentclass[11pt]{article}

\usepackage[english]{babel}
\usepackage{amsmath,amssymb,gensymb}
\usepackage{amsthm,amsfonts,dsfont,tensor,stmaryrd,manfnt,mathrsfs,svrsymbols,mathtools}
\usepackage[mathstyleoff]{breqn}
\usepackage{slashed}
\usepackage{enumerate}
\usepackage[normalem]{ulem}
\usepackage[mathscr]{euscript}
\usepackage{mathdots}
\usepackage{siunitx}
\usepackage{array}
\usepackage{multirow}
\usepackage{tabularx}
\usepackage{extarrows}
\usepackage{tabularx,booktabs}
\usepackage{enumitem}
\usepackage{colortbl}
\usepackage{xcolor}
\usepackage{sidecap}
\usepackage[font={small}]{caption}
\numberwithin{equation}{section}
\usepackage{soul}

\usepackage{tikz}
\newcommand*\circled[1]{\tikz[baseline=(char.base)]{
            \node[shape=circle,draw,inner sep=2pt] (char) {#1};}}

\usepackage{jheppub}
\usepackage{geometry}
\title{Navigating string theory field space with geometric flows}
\author[a]{Saskia Demulder,}
\author[b,c]{Dieter L\"ust,} 
\author[b]{Thomas Raml\,}

\affiliation[a]{Department of Physics, Ben-Gurion University of the Negev,\\ David Ben Gurion Boulevard 1, Beer Sheva 84105, Israel}
\affiliation[b]{Max-Planck-Institut f\"ur Physik (Werner-Heisenberg-Institut),\\ Boltzmannstr. 8, 85748, Garching, Germany}
\affiliation[c]{Arnold-Sommerfeld-Center for Theoretical Physics,\\
   Ludwig-Maximilian-Universit\"at, 80333 M\"unchen, Germany}

\abstract{
The Swampland Distance Conjecture postulates the emergence of an infinite tower of massless states when approaching infinite-distance points in moduli space. However, most string backgrounds are supported by fluxes, and therefore depart from the purely geometric paradigm. This fact requires an extension of the Swampland conjectures to scalar field spaces with non-trivial potentials, rather than just moduli spaces.  
To address this task, we utilise geometric flows, in particular generalised Ricci flow, to probe the associated scalar field spaces. Considering internal spaces supported by three-form fluxes, we first show that the distance defined in terms of the Perelman entropy functional needs to be refined in order to encompass fluxes. Doing so, we extend the Ricci Flow Conjecture to include  Kalb-Ramond flux besides the metric and the dilaton field. This allows us to probe infinite-distance points within these scalar field spaces in a purely geometric way.  We subsequently construct a geometric flow for internal manifolds supported by Ramond-Ramond fluxes and discuss its role in the Ricci Flow Conjecture. Our analysis suggests that in the presence of fluxes the Distance Conjecture might be better characterised in terms of a cost function on the space of metrics, rather than a genuine distance.}

\begin{document}

\vspace*{-1.5cm}
\begin{flushright}
  {\small
 MPP-2024-245\\
LMU-ASC 22/24
  }
\end{flushright}

\vspace{1.5cm}
\maketitle
\newpage

\section{Introduction} 

  The Swampland program \cite{Vafa:2005ui} aims to identify the effective field theories that can be consistently coupled to a theory of quantum gravity.  A central aspect of this program is the systematic exploration of all possible effective theories, testing their compatibility with a set of properties believed to be essential for a consistent theory of quantum gravity. One such criterion is the Swampland Distance Conjecture (SDC) \cite{Ooguri:2004zv}, which postulates that for viable theories, infinite-distance points in the moduli space of the effective theory imply the presence of an infinite tower of massless states. The infinite distances usually correspond to couplings or geometric parameters of the theory being sent to extreme values.   Generalising the idea of SDC, the Generalised Swampland Distance Conjecture \cite{Lust:2019zwm} applies the Distance Conjecture to the variations of the space-time metric itself, rather than scalar moduli. The relevant distance here is measured using the deWitt metric\footnote{In particular, following the notation of \cite{Headrick:2006ti} the deWitt$_{k}$ metric with $k=-1$.} \cite{DeWitt:1967yk},
\begin{align}\label{eq:deWitt}
	\Delta_g=c\int_{\lambda_i}^{\lambda_f}\mathrm d\lambda\left(\frac{1}{\mathrm{Vol}(M)}\int_{M}\sqrt{g}g^{MN}g^{OP}\frac{\partial g_{MO}}{\partial \lambda}\frac{\partial g_{NP}}{\partial \lambda}\right)^{1/2}\,,
\end{align}
where $\lambda$ parametrises the one-dimensional path, $g$ is a Riemannian metric on the manifold $M$, and $c$ is a dimensionful constant parameter of order one.  Varying the metric one effectively slowly evolves, in a usually smooth manner, from one geometry to the other.
Furthermore, going beyond moduli spaces, for which the conjecture was originally formulated,  
 recent efforts have increasingly concentrated on extending the conjecture to scalar field spaces with potentials  \cite{Calderon-Infante:2020dhm,Demulder:2023vlo,Basile:2023rvm,Mohseni:2024njl,Debusschere:2024rmi}. In the present paper we aim to make progress in this very direction, exploiting a natural way to explore the string theory field space.

Scanning through geometries is a familiar problem to mathematicians, more notably through so-called geometric flows. Given some geometrical data the space consisting of all such possible data, for example the space of Riemannian or K\"ahler metrics, can be explored by varying the initial manifold in a pre-assigned way. A celebrated example is Hamilton's Ricci flow \cite{Hamilton1982} and Perelman's combined flow \cite{perelman2002entropy,Perelman2003}. We first turn to the Ricci flow which consists of a set of partial differential equations driving the metric in a way very similar to a heat equation. Here, the Ricci tensor plays a role akin to the Laplacian of the heat equation, ``diffusing'' irregularities in the manifold's geometry, and thus smoothening out the metric over time. Ricci flow and the multitude of other geometric flows are pivotal mathematical tools to classify and explore geometries. For a given geometrical flow, a given initial manifold is evolved under the flow  towards canonical forms and thus revealing their topological and geometric properties.

Not surprisingly, geometric flows, and Ricci flow in particular, naturally emerge in physics, most notably in their relation to the renormalisation group equations \cite{Oliynyk:2005ak,Tseytlin:2006ak}.  In two-dimensional non-linear $\sigma$-models (NLSM) the one-loop renormalisation group (RG) flow corresponds to the Ricci flow of the target space metric. This relationship provides a geometric interpretation of RG flow by relating the evolution of coupling constants under scale transformations to the deformation of the manifold's geometry via Ricci flow.  More recently, in  \cite{Papadopoulos:2024uvi,Papadopoulos:2024tgs} the authors showed, how by differentiating between fixed points and solitonic solution the Ricci flow, scale invariance implies conformal invariance in a two-dimensional $\sigma$-model with a compact target space. In the context of black holes, the authors of \cite{DeBiasio:2022nsd} extended Ricci flow to the Einstein-Maxwell action, in order to analyse the stability of Reissner-Nordstr\"om black holes.

More relevant to the aims of this paper is the manner in which the Ricci flow  was used in \cite{Kehagias:2019akr}.  While Ricci flow in its original form only prescribes an evolution for the metric $g$, Perelman combined the flow of the metric with a flow equation for the dilaton\footnote{The relation between the scalar field introduced by Perelman and with dilaton in the Einstein-Hilbert action is rather subtle. We  discuss this point extensively in section \ref{sec:role_f}. For simplicity we identify the two in this introduction.} $\phi$. It is then  natural to think of the combined Perelman flow as defining paths in a generalised field space  
\begin{align}\label{eq:flow_simple}
   \mathbb R\rightarrow \mathcal{M}_g\times \mathcal{M}_\phi: \quad t\mapsto (g(t),\phi(t))\,,
\end{align}
where $\mathcal{M}_g$ denotes the space of Riemannian metrics on a given manifold $M$, $\mathcal{M}_\phi$ the corresponding space of dilaton configurations and $t$ is the flow parameter.
One of us and two other authors considered precisely this mechanism to vary the geometric data of the metric and dilaton and proposed that following Ricci flow towards a fixed point at infinite distance in the space of background metrics must be accompanied by an infinite tower of states \cite{Kehagias:2019akr}. This conjecture is dubbed Ricci Flow Conjecture (RFC) and was later applied in various contexts in the Swampland program \cite{DeBiasio:2020xkv,Velazquez:2022eco,DeBiasio:2022zuh,DeBiasio:2022omq}.

\begin{figure}[t]
    \centering
    \includegraphics[width=0.39\linewidth]{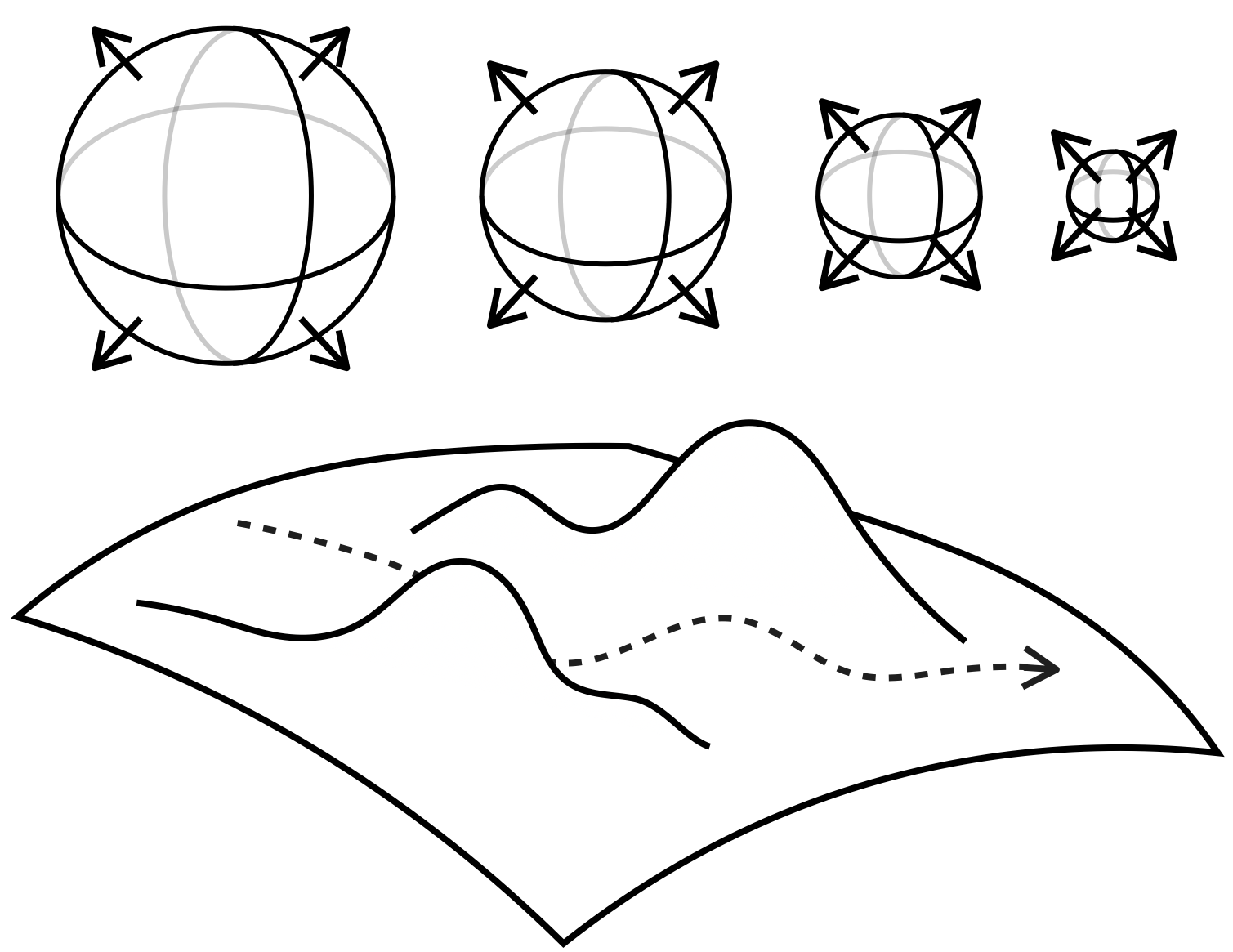}
    \caption{This figure  illustrates how the fluxes and curvature, rather than requiring direct extraction of the potential on scalar field space, govern the flow and the corresponding paths in scalar field space. As the flow evolves and modifies the geometry, the associated scalar fields dynamically adjust along a trajectory that mirrors the path followed in the scalar field space.}\label{fig:flows_global}
\end{figure}

In this paper our goal is to generalise the range of applicability of the Ricci Flow Conjecture to flux-supported internal geometries. We aim to study how, by allowing the internal space to vary under geometric flows, infinite distances in scalar field space can be accessed\footnote{We remark that our strategy here is somewhat orthogonal to some recent bottom-up approaches. While a clear disadvantage of our proposal is that not all infinite-distance points might be accessed using a geometric flow, it does however not require working directly with any potentials. In fact, having to consider the potential is circumvented entirely.}. Working at the level of the spacetime rather than the scalar field space has the distinct advantage of offering the mathematical ease to include fluxes into the analysis.  More precisely, we propose a different viewpoint to this issue by lifting the discussion from scalar field space back to the underlying geometrical picture before compactification, i.e. in some sense geometrising the abstract path or distance on scalar field space. By using geometric flows, we have no potential and we can use the chosen flow and its functional in order to quantify lengths. More precisely, our goal is to track infinite distances by varying the compact part of a string background and the flux supporting it. This approach is summarised in fig. \ref{fig:flows_global}.

This paper is structured as follows.
 In section \ref{sec:Ricci_flows}, we briefly review Hamilton's Ricci flow and the Swampland Ricci Flow Conjecture.  We then review a generalised geometric flow of \cite{garcia2021generalized} that includes NSNS-flux. Finally, we argue how geometric flows can probe the scalar field space, by effectively visiting different paths in that space as the geometry evolves along the flow.

In section \ref{sec:def_distance}, we define a measure of distance derived from measuring the length along the gradient of the entropy functional capturing a geometric flow. This more general distance measure reduces to the original distance measure proposed in \cite{Kehagias:2019akr}, thus offering a first-principle derivation. As the role of the dilaton is a critical ingredient in Perelman's combined flow, we show the implications of weakening the associated unit volume constraint. Using the three-sphere with H-flux as a case study one can easily re-derive, using generalised Ricci flow, the conclusions drawn in \cite{Demulder:2023vlo},  provided the original Ricci flow conjecture is altered. This modification, which is a direct consequence of the non-trivial fluxes, mirrors the behaviour induced by a potential on scalar field space. Finally, closing the loop we extend the analysis to the T-dual space.

In section \ref{sec:gen2_Ricci_flow}, we invoke new flow equations in order to capture the internal spaces of type IIA/B and 11D supergravity. We prove the existence of short-lived solutions, and discuss the monotonicity of the associated entropy functional. Finally, we illustrate the behaviour of  the  11D supergravity flow when for an internal four-sphere supported by RR-fields.

In section \ref{sec:distance_vs_cost}, collecting the results and insights derived from earlier sections, we scrutinise our proposed measure as a genuine mathematical notion of distance. Arguing in favour of cost measure, we then go on to contrast our conclusions with other recent endeavours related to probing scalar field space and extending the Swampland Distance Conjecture.

We conclude in section \ref{sec:conclusions}.

\section{The Ricci flow, generalised Ricci flow and the Ricci flow conjecture}\label{sec:Ricci_flows}

In the following section we review the necessary ingredients to tackle flux-supported internal manifolds. We first provide a review of the Ricci and Perelman flow, emphasising the properties relevant to the statement of the Swampland Ricci flow conjecture. In passing we highlight the relation between Ricci flow and renormalisation group flow. We then introduce an extension of the Ricci flow which naturally includes three-form flux besides the metric and the dilaton. 
With these elements at hand we lay out how more generally geometric flows can offer a systematic way to explore the space of metrics.

\subsection{Ricci flow and its Swampland conjecture}\label{sec:ricci_SDC}
We start by reviewing the Ricci flow and its associated conjecture. In absence of fluxes and dilaton one way to vary the metric in a natural way is to consider geometric flows, in particular Ricci flow:
\begin{align}\label{eq:Ricci_flow}
	\frac{\partial}{\partial t}g_{ij}(t)=-2R_{ij}\,,
\end{align}
where $R_{ij}$ is the Ricci tensor for $g_{ij}$ and $t$ is the flow parameter. Hamilton’s Ricci flow is a natural evolution of the metric that smooths out irregularities in the geometry over time. This process can be thought of as a gradient flow associated with the entropy functional introduced by Perelman: 
\begin{align}\label{eq:entropy_functional}
	\mathcal F(g,f)\equiv \int \, \mathrm{d}x^n \sqrt{|g|}\,
	e^{-f}\left(R+|\nabla f|^2 \right)\,.
\end{align}
where $f$ is a smooth real-valued function on $M$, which  is essentially the dilaton up to a factor: $f=2\phi$. Varying $\mathcal{F}(g,f)$ and under the additional constraint that the weighted integrated volume element is constant along the flow 
\begin{align}\label{eq:extra_cond_gradient_flow}
   \int \mathrm{d}x^n \sqrt{|g|}\,e^{-f}=1
\end{align}
the Ricci flow in eq. \eqref{eq:Ricci_flow} is gradient with respect to the ``Einstein Hilbert action'' in eq. \eqref{eq:entropy_functional} with the additional equation\footnote{After performing a diffeomorphism in order to decouple the two equations, cf. section \ref{sec:generalised_Ricci}.}
\begin{align}
	\frac{\partial g_{ij}}{\partial t}=-2 R_{ij}\,,\qquad \frac{\partial f}{\partial t}=-\Delta f +|\nabla f|^2-R\,.
\end{align}
In \cite{Kehagias:2019akr}, the authors proposed that following the gradient flow of background metrics toward a fixed point at infinite distance is associated with the emergence of an infinite tower of states in quantum gravity. In particular, the new conjecture proposes  that the entropy functionals of gradient flows can define a generalised distance in the space of background fields. This distance is measured by the logarithmic difference of the entropy functional 
\begin{align}\label{eq:Delta_F_standard}
    \Delta_\mathcal{F}=\log\left(\frac{\mathcal F_f}{\mathcal F_i} \right)\,.
\end{align}
For simple cases this distance is related to the scalar curvature of the manifold, expressed as
$\Delta \simeq \log R$. This proposal was further cemented by the retrieval of the AdS-Distance Conjecture, showing that the Ricci-time parameter $t$ is affinely related to the Weyl rescaling parameter.

Moreover, for constant metric and zero curvature, the flow of the function $f$ corresponds to the flow of the dilaton (see also the discussion in section \ref{sec:distance_non_gradient}). In this case the flow leads to weak string coupling, where a tower of
light string excitations is expected. In this way the entropy functional $\mathcal F(g,f)$ nicely comprises the two cases where one expects light states at infinite distance, as conjectured by the emergent string conjecture \cite{Lee:2019wij}, 
namely either light KK modes or light string states at weak string coupling.

\subsubsection{Flowing flux-supported internal spaces}

In this section we discuss how by starting from the Einstein Hilbert action and taking into account fluxes we are naturally led to consider extensions of the Ricci flow. This in turn guides us in the next section towards extending the Ricci flow conjecture. 

Consider first a total space which we assume to be a direct product space $M_{D}=M_{d}\times K_{n}$, where $M_{d}$ and $K_{n}$ are respectively the non-compact and compact part of the $D$-dimensional string background. Correspondingly, the background dynamics is determined by the Einstein-Hilbert action\footnote{Here and in the rest of the paper, for a $n$ index object $M_{i_1,\dots,i_n}$ we write $|M|^2$ for the contraction of all indices using the metric, i.e  $|M|^2=g^{i_1j_1}\dots g^{i_nj_n}M_{i_1\dots i_n}M_{j_1\dots j_n}$, which is the natural norm in this context.}
\begin{align}\label{eq:totEH}
    \mathcal{S}^{(D)}_\mathrm{EH}=\int\mathrm d ^D x\,\sqrt{|G|}\,e^{-2\Phi}\left(R(G)-\frac{1}{12}|H|^2+4|\nabla \Phi|^2\right)\,,
\end{align}
where $G$ and $\Phi$ are the metric and dilaton on the total space and $H$ will henceforth be assumed to only have legs in the internal space.
Usually, starting from this action, the next step is to compactify and study the resulting scalar field space and associated scalar potential. We will comment on this possibility in subsection \ref{sec:rel_dist}. In this paper however we take a different route.

Under the assumption that the background data respects the product structure, i.e. $G=\hat{g}\oplus g$ and $\Phi=\hat{\phi}+\phi$ the total Einstein Hilbert action in eq. \eqref{eq:totEH} ``factorises'' as
\begin{align}
	\mathcal{S}^{(D)}_\mathrm{EH}=\int\mathrm d ^D x\,\sqrt{|\hat{g}||g|}e^{-2\hat{\phi}-2\phi}\left[\left(R(\hat{g})+4|\nabla\hat{\phi}|^2\right)+\left(R(g)-\frac{1}{12}|H|^2+4|\nabla\phi|^2\right)\right]\,.
\end{align}
As we want to capture the data of the scalar field space, here we focus on the second term capturing the internal contribution. We then introduce\footnote{From now on we suppress the dependence on metric $g$ and write simply $R$ instead of $R(g)$ for the Ricci scalar and similar for other quantities.} the ``Einstein Hilbert functional'' capturing the internal geometry
\begin{align}
	\mathcal{S}_\mathrm{EH}&=\int\mathrm d ^n x\ \sqrt{|g|}e^{-2\phi}\left(R-\frac{1}{12}|H|^2+4|\nabla\phi|^2\right)\,.\label{eq:effective_action_compact_part}
\end{align}
  Remarkably, as we review in the next section, the functional in eq. \eqref{eq:effective_action_compact_part} serves as the basis to generate a gradient flow yielding the so-called \textit{generalised} Ricci flow.

\subsection{Generalised Ricci flow and scalar field space}\label{sec:generalised_Ricci}

Guided by a geometric interpretation of  the $\beta$-functions a natural extension of the Ricci flow equation was obtained in \cite{Oliynyk:2005ak}, see also \cite{streets2008regularity}. This flow, later dubbed ``generalised Ricci flow'', naturally extends the Ricci flow to include the contribution from a Kalb-Ramond field. Realising its natural implementation into generalised geometry \cite{Garcia-Fernandez:2016ofz,streets2017generalized}, a fact which is shortly reviewed in appendix \ref{app:GG_and_T-duality_form}, this flow has been extensively studied. We refer to the dedicated book \cite{garcia2021generalized} for an extensive introduction and exhaustive bibliography.

In \cite{Oliynyk:2005ak} it was shown how the flow defined  by the $\beta$-functions, following the blueprint identified by Perelman for the Ricci flow, is a gradient flow for for the Einstein-Hilbert action. That is, the new \textit{generating functional} of interest is the \textit{generalised Perelman functional} given by
\begin{align}\label{eq:SEH}
   \mathcal F(g,H,f)=\int_M d V_g\,e^{-f}\left(R-\frac{1}{12}|H|^2+|\nabla f|^2 \right)\,,
\end{align}
where we defined $d V_g= \mathrm{d}x^n \sqrt{|g|}$.
One can readily recognise this as the Einstein-Hilbert functional $\mathcal{S}_{\mathrm{EH}}$ with the Kalb-Ramond flux $H$ turned on. The crucial observation is that the variation of this functional together with the same constant volume condition from above
\begin{align}\label{eq:volumeconstr}
    \lambda(g,H)=\inf_{\{f\,\mid\,\int_Me^{-f}\mathrm d V_g=1\}}\mathcal F(g,H,f)\,,
\end{align}
leads again to a gradient flow \cite{garcia2021generalized}
\begin{gather}
\begin{aligned}\label{eq:gRicci_flow_with_f}
    \frac{\partial}{\partial t}g_{ij}&=-2R_{ij}+\frac{1}{2}H_{ij}^2 -2\nabla_i \nabla_j f\,,\\
    \frac{\partial}{\partial t}B_{ij}&=\nabla^k H_{kij} - H_{kij}\nabla^k f\,,\\
    \frac{\partial}{\partial t}f&=- R+\frac{1}{4}|H|^2-\Delta f\,.
\end{aligned}
\end{gather}
This is known as \textit{generalised Perelman flow} or \textit{generalised combined flow} and from now on whenever we write $\mathcal{F}$, we implicitly assume that the constraint \eqref{eq:volumeconstr} is imposed, while we use $\mathcal{S}_{\mathrm{EH}}$ for the functional without any constraint imposed.

The right hand side of the flow equations are now nothing else than equations of motions of the action $\mathcal{S}_{\mathrm{EH}}$ or equivalently the $\beta$-functions of the associated NLSM with background data $(g,B)$, i.e. $\partial_t g = -2 \beta^g, \partial_t B = -2 \beta^B$. The interpretation for $f$ is less straightforward. In fact the flow equation for $f$ follows from the constraint in \eqref{eq:volumeconstr} and is not the equations of motion or $\beta$-function of the standard string theory dilaton upon setting $f=2\phi$. However, whenever we are at a fixed point of the flow of $(g,b)$, the flow equation of $f$ reduces to $\beta_\phi$ and therefore indeed at the end of the flow we can identify $f=2\phi$ with a on-shell dilaton.

A notable feature of Ricci flow is its invariance under diffeomorphisms; that is, if  $\psi_t$  is a one-parameter family of diffeomorphisms, then the pullback metrics  $\psi_t^* g(t)$  also satisfy the Ricci flow equation. This implies in particular that the Ricci flow is not strictly parabolic, which complicates the analysis of its behaviour. We will return in detail to this issue in a more general instance of the flow in section \ref{sec:flow_typeII}. For the time being, this invariance property enables to perform a diffeomorphism $\psi$ generated by the vector field $\nabla f$. The system \eqref{eq:gRicci_flow_with_f} is then brought into the form
\begin{gather}
\begin{aligned}\label{eq:gRicci_flow}
    \frac{\partial}{\partial t}g_{ij}&=-2R_{ij}+\frac{1}{2}H_{ij}^2\,, \\
    \frac{\partial}{\partial t}B_{ij}&=\nabla^k H_{kij}\,,\\
    \frac{\partial}{\partial t}f&=- R+\frac{1}{4}|H|^2-\Delta f+ (\nabla f)^2\,,
\end{aligned}
\end{gather}
therefore decoupling $f$ from the other flow equations. This form of the generalised combined flow is known as \textit{gauge fixed generalised (Perelman) flow} and to the system without the flow equation for $f$ simply as \textit{generalised Ricci flow}. In the remainder of this work we will most of the time loosely refer to all the different nuances of the flow simply as generalised Ricci flow. Note that upon integration by parts and using $\Delta(e^{-f})=e^{f}|\nabla f|^2-e^{-f}\nabla f$ one can write $\mathcal F$ in the form

It is worth noting that fixed points of the generalised Perelman flow coincide with the critical points of the functional $\mathcal{F}$. However when only considering generalised Ricci flow, i.e. when discarding the dilaton and its associated flow, fixed points of the flow will in general not correspond to critical points of the Einstein-Hilbert functional. In order to understand this, note that an extremal value of $\mathcal{S}_{\mathrm{EH}}$ is reached, see proposition 3.46 of \cite{garcia2021generalized}, if and only if 
\begin{align}\label{eq:iff_extremal_EH}
	R^+_{ij}-\tfrac{1}{2}R\,g_{ij}-\tfrac{1}{12}|H|^2g_{ij}=0\,,
\end{align}
where $R^+_{ij}$ is the Bismut-Ricci tensor defined in eq. \eqref{eq:BismutRicci}.
This is equivalent to asking for the manifold to be Bismut flat and have a vanishing generalised scalar curvature
\begin{align}\label{eq:critpt}
    R^+=0\,,\qquad S=0\,,
\end{align}
where $S$ is the generalised scalar curvature defined in eq. \eqref{eq:scalarcurv}.
The condition for a critical point \eqref{eq:critpt} is clearly stronger than just asking for a fixed point of the generalised Ricci flow, i.e. $\partial_t g=0$ in \eqref{eq:gRicci_flow}. 
In particular, and as expected, while for the conventional Ricci flow fixed points and extremal points of the Einstein-Hilbert action are one and the same thing, this is no longer true for the generalised flow when discarding the scalar field. 

However when working with $\mathcal{F}$ and the associated generalised Perelman flow, the condition of having $S=0$ in addition to Bismut flatness is trivialised and therefore fixed points of this flow are indeed critical points of its generating functional.

Furthermore, let us mention that due to the fact that the flow equations coincide precisely with the renormalisation group flow   $\beta$-functions of the non-linear $\sigma$-model with background $(g,B)$ one can try to give an interpretation of Ricci flow in terms of an RG flow. We comment on this further in appendix \ref{app:RG-flow}.

\subsection{Simple examples of the generalised Ricci flow} \label{sec:gRicci_simple_examples}
Let us first briefly review the simplest solution to gain some intuition for the Ricci and the generalised Ricci flow: Einstein manifolds, whose  Ricci curvature takes the form
\begin{align}\label{eq:Einstein_m}
	R_{ij}(g)=k g_{ij}\,,
\end{align}
for $k$ a real constant. Within this classes of manifolds we illustrate and contrasts the two flows introduces in this section through three examples: the three-sphere with and without flux and hyperbolic space with flux. 

\paragraph{Three-sphere without flux.} A particular example is the $n$-sphere of radius $r$ for which $k$ in eq. \eqref{eq:Einstein_m} is positive with $k=(n-1)/r^2$.  
For later purposes it is illustrative to write the flow equation in terms of the radius $r$, which will become time dependent under the flow. Indeed taking the ansatz $g(t)=r^2(t)g_0$ with $g_0$ the metric of the round unit sphere, the conventional Ricci flow equation reads
\begin{align}\label{eq:flow_sphere}
\frac{\partial}{\partial t}r(t) = - \frac{(n-1)}{r(t)}\,,
\end{align}
such that the radius of the sphere shrinks to zero as
\begin{align}
    r(t) = \sqrt{r_0-(n-1)t}\,.
\end{align}

\paragraph{Three-sphere with $H$-flux.} Now the simplest example of generalised Ricci flow is to support the sphere with $k$ units of $H$-flux. In particular in three dimension there is just one possible choice of $H$-flux as it necessarily is proportional to the volume form
\begin{align}
	H=2h\,\mathrm d\mathrm{vol}\,,
\end{align}
where $d\mathrm{vol}$ is the volume form of the unit three-sphere and $h$ is a real number. Explicitly, one can easily show\footnote{This follows directly from computing $R^+$, observing it only possess a symmetric part and applying lemma 4.6 of \cite{garcia2021generalized}.} that $\partial_t h=0$. The flow equation reduce to a differential equation for $r(t)$
\begin{align}
	\frac{\partial}{\partial t}r(t) = -\frac{2}{r(t)}+\frac{2h^2}{r(t)^4}\,.
\end{align}
Taking $h=0$ we are back to standard Ricci-flown sphere which collapses after a time $T$. The more interesting solution is when $h\neq 0$, so we deal with a genuine generalised Ricci flow and in this case the differential equation admits a (unique) fixed point: $r=\sqrt{h}$.  This fixed point is in fact attractive since when $r<\sqrt{k}$ it follows that $\partial_t r(t)>0$ and when $r>\sqrt{h}$ then $\partial_t r(t)<0$. For $t\rightarrow \infty$ the solution asymptotes to $r(t) \rightarrow \sqrt{h}$.
We see that the $H$-flux has removed the singularity that would be present in the conventional Ricci flow, effectively rendering the solution convergent at infinity.

We conclude that the three-sphere with $H$-flux, which realises the background of the $SU(2)$-WZW model at level $k=h$ correlated to the radius of the sphere via $r=\sqrt{h}$, is Bismut flat, and thus a fixed point of the flow. This is a  critical new feature introduced when fluxes are included to the Ricci flow equation: while in the original flow the geometry may collapse into a singularity, those will now be stabilised and circumvented.

This provides a first hint that the Ricci Flow Conjecture in the presence of $H$-flux is in fact richer than the standard one. Indeed, although this geometry is a fixed point, we do not expect any set of massless states for a sphere fixed at $r=\sqrt{h}$.  We show in section \ref{sec:potentials_and_flow} that this intuition neatly mirrors the results of \cite{Demulder:2023vlo}: the geometry at $r=\sqrt{h}$ does not lie at infinite distance.

\begin{figure}[t]
	\centering
	\includegraphics[scale=0.69]{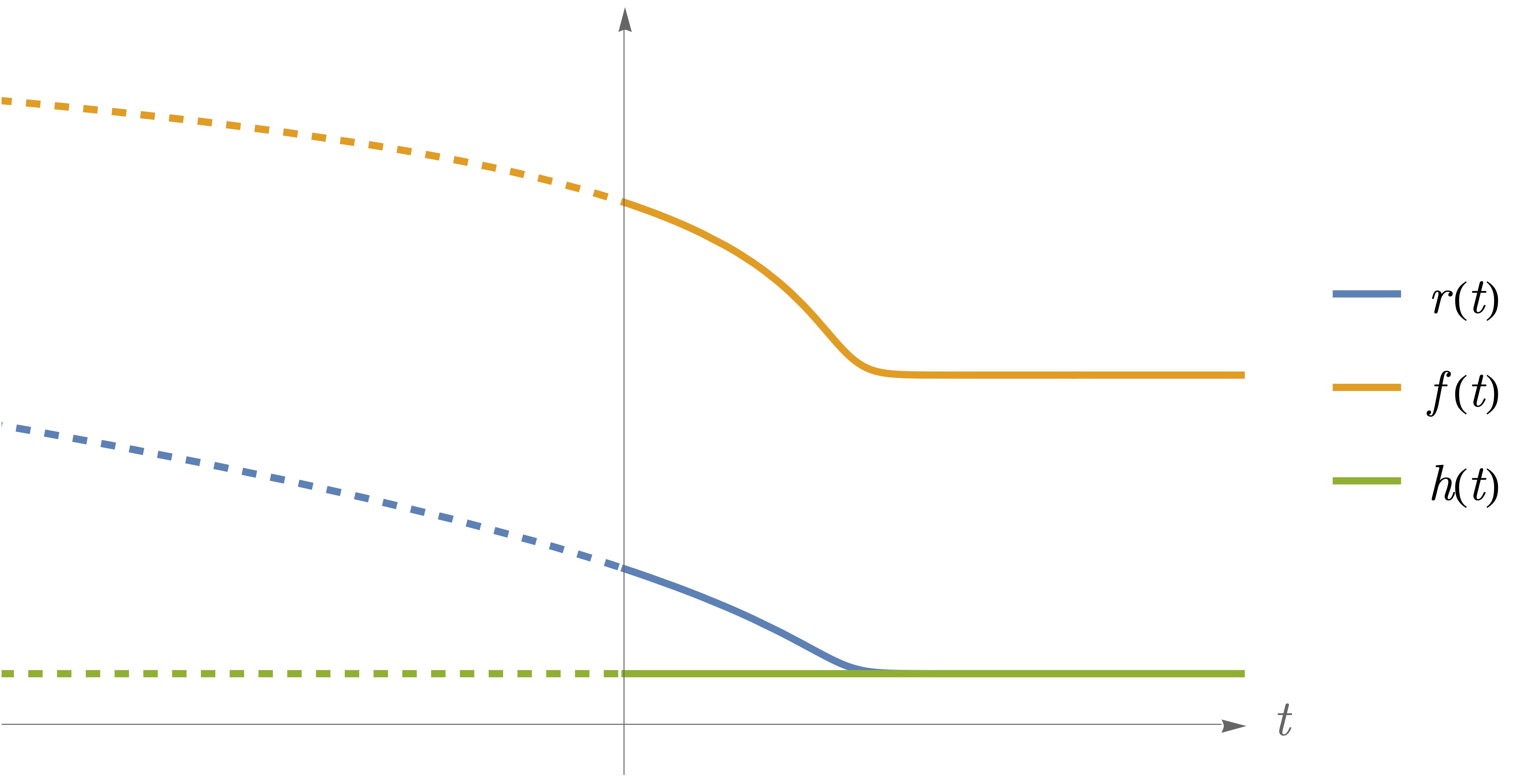}
	\caption{Plotted is the flow behaviour of the combined flow of the three-sphere with $h=1$ units of $H$-flux starting at an initial value of $r(t_0)> \sqrt{h}$ (solid lines). For $t \to \infty$ the radius $r(t)$ (blue) decreases and stabilised at the fixed point $\sqrt{h}=1$. The flux number $h$ (green) is constant in time and the scalar $f$ (orange) stabilises in a way such that all along the flow the generalised volume element stays constant. The flows can be extended backwards in time from $t_0$ (dashed lines).}
    \label{fig:S3_H_with_f_backwards}
\end{figure}

We finally also state the full solution of the generalised combined flow, including the scalar field $f$
\begin{align}
\frac{\partial}{\partial t}r(t) = -\frac{2}{r(t)}+\frac{2h^2}{r(t)^4}\,,\qquad
\frac{\partial}{\partial t}h(t) =0\,,\qquad\frac{\partial}{\partial t}f(t) = -\frac{6}{r(t)^2}+\frac{6 h^2}{r(t)^6}\,.
\end{align}
Note that since $f$ is only a function of $t$ and not the coordinates, the gauge fixed version of the flow and the one obtained directly from the functional coincide. We see that similar to $r(t)$, $f(t)$ flows to a finite fixed point. It is easy to check that indeed the weighted volume element is constant. 
The flow behaviour is illustrated in figure \ref{fig:S3_H_with_f_backwards}.

\paragraph{Three-dimensional hyperbolic space.}
When on the other  hand we choose an Einstein manifold with  $k$ negative, i.e. a compact hyperbolic 3-manifold, $R_{ij}(g)=-2g_{ij}$ such that its radius behaves as
\begin{align}
	\frac{\partial}{\partial t}r(t) = \frac{2}{r(t)}\,,
\end{align}
and the manifolds expands (homothetically) forever. Including again a flux proportional to the volume form one obtains \cite{garcia2021generalized}
\begin{align}
	\frac{\partial}{\partial t}r(t) = \frac{2}{r(t)}+\frac{2h^2}{r(t)^4}\,.
\end{align}
Therefore we see that in this example in contrast to the sphere the addition of flux does not render the flow convergent for long flow times. The behaviour for large $r(t)$ stays essentially unchanged.

\subsection{Flowing through the space of metrics and scalar field space}

Following the generalised Ricci flow of a given initial data $(g_{ij},B_{ij},f)$, leads to a one-parameter trajectory through the space of (generalised) Riemannian metrics on the associated manifold $M$, denoted $\mathcal{GM}$:
\begin{align}\label{eq:flow_wH}
   \mathbb R\rightarrow \mathcal{GM}\times \mathcal{M}_f \cong \mathcal{M}_g \times \Omega^2(M)\times  \mathcal{M}_f:  \quad t\mapsto (\mathcal{G}(t),f(t))\,,
\end{align}
where each point on along the flow is a generalised metric $\mathcal{G}(g,B)$, defined in eq. \eqref{eq:gen_m}, or a point in $\mathcal{GM}$.  This directly generalises the purely geometric scenario in eq. \eqref{eq:flow_simple}, where only the metric and dilaton (or $f$) were considered. Note that here again we encounter the subtle difference of differentiating between the dilaton $\phi$ and the scalar $f$, a point which we will return to in section \ref{sec:role_f}.

The (generalised) Ricci flow leading to the map in eq. \eqref{eq:flow_wH} -- in fact any geometric flow for that matter -- gives a means to probe the space of (generalised) metrics. In the present work we propose to use Ricci flow, or more generally geometric flows, in order to make progress in the exploration of scalar field spaces arising from compactification and therefore go beyond the moduli space paradigm in the SDC. The main observation is that by utilising geometric flow there is no need to study the scalar field space directly, extracting its metric and potential which is sourced by curvature and fluxes. Instead, considering flows we lift the discussion from the scalar field space to a purely geometric enquiry: solving the flow equations in terms of the relevant curvature and flux terms characterising the solution.

This perspective offers the distinct advantage of alleviating the need to directly extract and work with the potential on scalar field space. Instead the fluxes and curvature that would otherwise give rise to this potential now directly control the flow and thus the paths visited in the scalar field space. Whilst the flow gradually modifies the geometry, the associated scalar fields change accordingly. More concretely, given an initial geometry the flow follows a trajectory which is mirrored by a path in scalar field space, see fig. \ref{fig:flows_global}.

Our proposal can only work if we can construct a distance-measure on the space of (generalised) Riemannian metrics which simultaneously accounts faithfully for the distance on scalar field space. In addition we need to single out a path in this generalised field space and translate it to the scalar field space. In the next section we show that this can be indeed realised. In particular, the flow selects a path that, when working with Perelman's combined flow, is gradient with respect to the associated action.
The metric on the generalised field space is the deWitt metric, which is the natural metric on this space. In fact we show that this also carries over for the generalised case of including the Kalb-Ramond $B$-field, leading to the Candelas-de la Ossa metric \cite{Candelas:1990pi}.

\section{Measures of distance along the flow}\label{sec:def_distance}
A crucial ingredient for formulating a generalised Ricci Flow Conjecture is a well-defined notion of distance (though whether this is a genuine measure of distance is discussed in section \ref{sec:distance_vs_cost}). In the case of standard Ricci flow, several possible definitions were proposed in \cite{Kehagias:2019akr}. In this section we explore the natural generalisations in the presence of flux and show that there is a more fundamental definition of distance whenever we have a generating functional at our disposal. For special cases, this new definition reduces to the known simple form given in \eqref{eq:Delta_F_standard} while generically also reproducing the generalised deWitt metric  in the presence of $H$ flux, introduced in \cite{Candelas:1990pi}.

In the previous sections we discussed the generalised Perelman functional $\mathcal{F}(g,H,f)$ with respect to which the generalised Ricci flow is gradient. Recalling therefore the proposed definition of distance along the combined flow of metric and scalar field $f$ \cite{Kehagias:2019akr}
\begin{align}
    \Delta_\mathcal{F} = \log \left(\frac{\mathcal{F}_{t_f}(g,f)}{\mathcal{F}_{t_i}(g,f)}\right)\,,
\end{align}
there is a natural generalisation in the presence of $H$-flux by simply replacing $\mathcal{F}(g,f)$ by $\mathcal{F}(g,H,f)$. 
However, a drawback of this approach is that apart from special cases it is not clear how to connect this notion of distance to the ``standard'' deWitt metric. Therefore  it would be preferable to define a more general notion of distance along the flow, that incorporates all of the above mentioned notions of distance. Indeed, we can define such a notion of distance by taking inspiration from the well-known line integrals along vector fields.

\subsection{A universal length from geometric flows}\label{sec:univLength}

Given a geometric flow, we have all the necessary tools to explore different trajectories through the space of (generalised) Riemannian metrics. In order to formulate the Distance Conjecture, one is left with identifying an adequate notion of distance or length between two points along the flow. Guided by the analogy between the Gibbons-Hawking entropy and Perelman's entropy, the distance $\Delta_{\mathcal F} \simeq \log \mathcal F$, where $\mathcal F$ is the Perelman functional in eq. \eqref{eq:entropy_functional} was defined in \cite{Kehagias:2019akr}. It was shown by Perelman in \cite{perelman2002entropy} as a central component of his proof that the function $\mathcal F$ is strictly monotonous along the flow. In the present setting this ensures that $\Delta_{\mathcal F}$ verifies at least one of the axioms for a distance. This definition was further cemented by matching with distance obtained from the AdS-Conjecture. However a first principle derivation for the distance $\Delta_{\mathcal F}$ was lacking.

In this section we show how, starting from a very natural definition for the length, we are lead to length measure $\Delta_L$. Specialising this new definition to the cases where the manifold is fluxless considered  in \cite{Kehagias:2019akr}, consistently reduces to $\Delta_L \simeq \Delta_{\mathcal F} \simeq \log \mathcal F$. Furthermore, we explicitly relate this new length $\Delta_L$ to the generalised deWitt metric \cite{Candelas:1990pi,Demulder:2023vlo} establishing the fact that, at least for gradient flows, we are in fact measuring distances according to the natural metric on this space. Ricci flow therefore mostly serves as a tool in defining the natural trajectories in this space of theories. In particular, although describing off-shell configurations along the flow, the paths follow the gradient of the natural functional in this setup, namely the string effective action. At the fixed points of the flow the geometries are on-shell. This approach, in particular, offers a natural mechanism for obtaining a controlled off-shell path that reverts to being on-shell precisely at the fixed point of the flow.

In order to start we remind ourselves of the simple notion of a line integral. For a given vector field  $\pmb{F}(x)$ and $\gamma$ a curve in $K_n$ the line integral along $\gamma$ is defined as
\begin{align}
    I_\gamma = \int_\gamma \pmb{F} (\pmb{r}(t)) \cdot \dot{\pmb{r}}(t) \ \mathrm{d}t\,,
\end{align}
with $\pmb{r}(t)$ a parametrisation of the curve $\gamma$. We now define the vector field $\pmb{F}$ as the gradient of the functional $\mathcal{F}$\footnote{The gradient is with respect to the basis 
\begin{align*}
    \left(
       \frac{\partial}{\partial g_{ij}} ,\frac{\partial}{\partial B_{ij}} \right)^T\,.
\end{align*}}
\begin{align}\label{eq:F_vector_field}
    \pmb{F}= \pmb{\nabla} \mathcal{F} \equiv \int_K \mathrm{d}V_ge^{-f}\Bigl((\beta_g)^{ij} \pmb{\partial}_{g_{ij}}+(\beta_B)^{ij} \pmb{\partial}_{B_{ij}}\Bigr)\,,
\end{align}
where the coefficients $\beta^{g/B}$ are essentially the flow equations given in \eqref{eq:gRicci_flow_with_f}, scaled by a factor of $-1/2$. 
Now inspired by the line integral we define\footnote{The factor $V_K$ is constant with respect to the flow time $t$ for the generalised Perelman functional due to the constraint. Therefore we could immediately  pull it outside the integral, however we want to make the notion of length also applicable even for more general functionals where the volume might not be constant in $t$, cf. section \ref{sec:distance_non_gradient}.}
\begin{align}\label{eq:Delta_L}
    \Delta_L = \int_t \tfrac{1}{\sqrt{V_K}} \sqrt{|\pmb{F}(\pmb{r}(t))\cdot \dot{\pmb{r}}(t)|}\, \mathrm{d} t=\tfrac{1}{\sqrt{V_K}} \int_t   |\partial_t \mathcal{F}|^{1/2}\, \mathrm{d} t\,,
\end{align}
where $V_K=\int_K \mathrm{d}V_g e^{-f}$ and $\pmb{r}(t) = (g_{ij}(t),B_{ij}(t))^T$. The square root was introduced in order to ultimately match with the already known measures of distance. 
On the other hand we can write 
\begin{align}\label{eq:Delta_L_v2}
    \Delta_L &=\tfrac{1}{\sqrt{V_K}}\int_t\,  \mathrm{d} t \left|\int_K \mathrm d V_g e^{-f} \Bigl((\beta_g)^{ij}\partial_t g_{ij} + (\beta_B)^{ij}\partial_t B_{ij}\Bigr) \right|^{1/2}\,  \nonumber\\
    &=\tfrac{1}{\sqrt{V_K}} \int_t\,  \mathrm{d} t \left|\int_K \mathrm d V_g e^{-f} \left(\tfrac{1}{2} g^{il}g^{jm}\partial_t g_{lm}\partial_t g_{ij} + \tfrac{1}{2}g^{il}g^{jm}\partial_t B_{lm}\partial_t B_{ij}\right) \right|^{1/2}  \nonumber\\
    &=\tfrac{1}{\sqrt{2 V_K}} \int_t  \mathrm{d} t\,\left|\int_K \mathrm d V_g e^{-f} \Bigl(\left|g^{-1}\partial_t g\right|^2+\left|g^{-1}\partial_tB\right|^2\Bigr) \right|^{1/2}\,,
\end{align}
where in the second line we used the definition of the generalised Ricci flow \eqref{eq:gRicci_flow}. 
We recognise the last line as the generalised deWitt metric which was encountered in \cite{Demulder:2023vlo}, evaluated along the path $\gamma$ selected by Ricci flow. This last expression  was already found and studied in \cite{Candelas:1990pi} as a natural metric on the space of parameters of  K\"ahler manifolds supported by a non-trivial $B$-field. Therefore as we argued in \cite{Demulder:2023vlo} and was also discussed  in \cite{Basile:2023rvm,Li:2023gtt, Shiu:2023bay,Palti:2024voy} from different perspectives, flux contributions enter the notion of distance in a non-trivial way and in particular cannot be ignored in general. Furthermore, note that the expression in the last line of \eqref{eq:Delta_L_v2} is a sum of positive terms and zero only at the fixed points, a fact which we come back to when discussing the properties of the distance in section \ref{sec:distance_vs_cost}.

In order to connect this notion of distance to the existing distances for Ricci flow we restrict to the case of conventional Perelman flow with just metric and scalar field $f$. In this case, the above last line is of course just the deWitt metric. However we can also consider the expression in eq. \eqref{eq:Delta_L_v2} which in this case reduces to\footnote{Time-independence of the volume element is again crucial for pulling the time derivative into the integral.}
\begin{align}
    \Delta_L  =\tfrac{1}{\sqrt{V_K}}  \int_t  \sqrt{|\partial_t \mathcal{F}|}\ \mathrm{d} t =\tfrac{1}{\sqrt{V_K}}  \int_t\, \mathrm{d} t\left|\int_K \mathrm d V e^{-f}\  \partial_t R(t)\right|^{1/2}\,.
\end{align}
If we further restrict to spaces with constant (in $x^\mu$) Ricci curvature such that $R_{\mu\nu} R^{\mu\nu}= \frac{k}{2} R^2$ 
and $k$ some constant it follows that $\partial_t R = k R^2$, such that
\begin{align}
    \Delta_L   &= \tfrac{1}{\sqrt{k}} \int_t \left| \frac{     \partial_t R(t)}{R(t)} \right| \mathrm{d} t \simeq \log\left(\tfrac{R(t_f)}{R(t_i)}\right)
    \simeq \log \left(\tfrac{\mathcal{F}(g,f)|_{t_f}}{\mathcal{F}(g,f)|_{t_i}} \right)\,.
\end{align}
Hence we recover the proposed distance $\Delta_{\mathcal{F}}$ in \cite{Kehagias:2019akr}.

Of course a similar computation shows that under certain conditions on  the metric $g$ and three-form flux $H$, the distance reduces to 
\begin{align}
    \Delta_{L} \simeq \log \left(\tfrac{\mathcal{F}(g,H,f)|_{t_f}}{\mathcal{F}(g,H,f)|_{t_i}}\right)
\end{align}
In the above, we worked with Perelman's Ricci flow, i.e. we assumed that the constant volume condition along the flow is fulfilled. Otherwise the normalising factor $V_K^{-1/2}$ is time-dependent and needs to be integrated over. Furthermore the gradient of $\mathcal{F}$ would no longer give the generalised flow. Below we investigate ways in which one can relax the unit volume constraints. These modifications however come at the price of losing monotonicity of the functional along the flow.  We explain that one can regardless obtain a well-defined notion of distance, this time using (a modified version) of the distance $\Delta_L$ above, while the naive choice $\log(\mathcal{F}_f/\mathcal{F}_i)$ turns out to be problematic as we explain. We will return to the interpretation of the measure $\Delta_L$ (or lack thereof) as a distance in section \ref{sec:distance_vs_cost}.

\subsection{Role of $f$ as a dilaton}\label{sec:role_f}\label{sec:distance_non_gradient}

In the previous section we introduced the notion of distance $\Delta_L$ in terms of generalised Perelman's functional $\mathcal{F}$ and therefore with the generalised unit volume constraint. As detailed in appendix \ref{app:modified_flows}, the unit volume constraint effectively eliminates a contribution from the variation of $\mathcal{F}$ that is proportional to the trace of $g$, by imposing a suitable flow equation for $f$. As such, the unit volume constraint is critical in obtaining a well-behaved flow. Physically the interpretation of this constraint appears somewhat artificial, and one could wonder how crucial it is in the context of the Swampland Ricci Flow Conjecture. In particular, it is only when the relation between the volume of the manifold and the dilaton are decoupled that one could contemplate accessing infinite distance associated with the string coupling while  at the same time  keeping the spatial volume finite. In \cite{Kehagias:2019akr}, these infinite distance points could be attained by considering the purely dilatonic flow. In this section, we discuss two ways in which the dilaton can be decoupled from the constant volume constraint, while preserving (some) of the desirable properties of the distance.

To address this question in this section, we explore the possibility of discarding the unit volume constraint and discuss the resulting flows. In the absence of the unity volume constraint, the additional term in the variation of the functional in eq. \eqref{eq:term_from_variation} has to be taken care of by a different mechanism. One could attempt to absorb the terms proportional to the trace part of $g$ into the corresponding flow equation for the metric $g_{ij}$ and therefore define a flow equation that is given by the Einstein field equation. It is well known that this ``Einstein flow'' is highly problematic flow and generically does not admit even short time solutions, which in fact can be seen as a manifestation of the conformal mode problem of gravity. We refer to \cite{Papadopoulos:2024uvi} for an insightful discussion. 

Instead we modify the flow equations at the expense of losing the gradient property. Indeed, as noted in \cite{garcia2021generalized}, there is the ``curiosity'' that the action $\mathcal{S}_{EH}$ or equivalently the generalised Perelman functional \textit{without} the unit volume constraint can remain a monotonous function along certain modified versions of the flow. However they are no longer acting as a generating functional for the flow equations. In particular, the resulting flows will not be gradient with respect to action $\mathcal{S}_{EH}$. 

Leaving the details of the derivation to appendix \ref{sec:modified_flows}, we obtain a discrete  family of flows labelled by an integer $\alpha$,
\begin{align}\label{eq:alpha_flow}
    \frac{\partial}{\partial t} f = -(\alpha + 1) R + \frac{\alpha+3}{12} |H|^2 - (2\alpha+1)\Delta f + \alpha |\nabla f|^2 \,.
\end{align}
From this expression we single out two immediate special cases. 
Choosing first $\alpha=0$ gives back the constrained flow. Choosing the parameter $\alpha$ instead to be strictly positive, the functional $\mathcal{S}_{EH}$ provides a monotonous function along the resulting flow which, however, is no longer gradient. For $\alpha=-1$ the resulting flow equation for $f$ turns out to be exactly the $\beta$-function of the dilaton. However, it comes at the expense of even losing the property of monotonicity of $\mathcal{S}_{EH}$ along the corresponding flow. We contrast the resulting flow equations and their potential physical interpretation in the appendix \ref{app:modified_flows}, as they will not play a major role for the main discussion to follow. 

Lastly we mention that there is in fact a possibility for explicitly introducing a dilaton $\phi$ as pointed out in \cite{streets2023scalar}. Its flow equation is the associated $\beta$-function, i.e $\partial_t \phi = -2\beta^\phi$ and the flow is gradient. The complete system is then described by $\{g,B,\phi\}$ all evolving according to their $\beta$-functions together with a scalar field $\tilde
f$ that takes the same role as $f$ for the combined flow.  However the gradient property is in the weaker sense that only the flow equations of $g$ and $B$ arise as directly form the generating functional, while the flow equation for $\phi$ has to be imposed by hand through a -- somewhat artificial -- splitting $f=2\phi+\tilde{f}$. The resulting flow successfully decouple the dilaton $\phi$ from the volume of the manifold. The details and explicit examples can be found in appendix \ref{app:modified_flows}.

While this discussion clearly shows that, despite weakening the unit volume constraint, such flows persist and one can decouple the probe points where the dilaton diverge, its use within the Swampland Ricci Flow Conjecture is far less clear. Indeed, the above modified versions of the flow come with different drawbacks as compared to the generalised Ricci flow that is obtained as a gradient from $\mathcal{F}$. 

In light of a later discussion of the properties of the distance $\Delta_L$, we further comment on the monotonicity of $\mathcal{F}$ for the modified flows in eq. \eqref{eq:alpha_flow}. We can again distinguish between two cases. 

Restricting first to values $\alpha \geq 0$, we need to ascertain if at any point of the flow $\partial_t \mathcal F=0$. This happens if and only if all the summands in the functional are identically zero. This in turn, as is explained in appendix \ref{app:modified_flows}, is only true at fixed points. We can thus conclude that the functional is strictly monotonous.

 If we would allow for $\alpha=-1$, the summands in the expression for $\partial_t \mathcal F$ can cancel each other such that $\partial_t \mathcal{F}=0$ even at points which are not fixed points. This is potentially problematic as such a situation can lead to a violation of the axiom of strict positivity, i.e. $\Delta(x,y)=0$ for $x\neq y$. We discuss this further in section \ref{sec:distance_vs_cost}.
We also stress again that whenever the unit volume constraint is not imposed for all $t$ the factor $1/\sqrt{V_K}$ can not be pulled out of the integral in $\Delta_L$.

\paragraph{Remark.}  
\begin{itemize}
	\item 
In \cite{Kehagias:2022mik}, the authors proposed an interesting interpretation for the unit volume constraint. They observed that one can obtain the equations of motion for the metric, Kalb-Ramond field and dilaton by using, in lieu of Einstein Hilbert action, the Perelman's functional together with the unit volume constraint. This is a known result, but their interpretation suggests that this principle does deeper than a mathematical curiosity. Implementing this idea they extended to incorporate higher order string-loop corrections, Ramond-Ramond forms and moduli fields.
\end{itemize}

\subsection{Revisiting the SDC and  three sphere with $H$-flux}\label{sec:potentials_and_flow}

As we have outlined in the introduction, one of the main motivation of this work is to obtain an alternative viewpoint on the role of potentials on scalar field space using generalised Ricci flow.  When considering fluxes, the resulting scalar field space will display a potential which besides extrema will also generically feature divergent directions. A simple (toy) example for such behaviour is provided by the $SU(2)$ WZW model. In \cite{Demulder:2023vlo}, we argued that those divergent points of the potential in the scalar field space go hand in hand with ``missing'' infinite towers of states, effectively removing what naively would be an infinite distance point. In the example of the three-sphere with $H$-flux this precisely occurs when taking the small radius limit. The three-sphere doesn't support winding modes and accordingly there is no tower of winding states that becomes massless in this limit. The curvature of the sphere and the non-trivial $H$-flux however generate a potential for the scalar corresponding to the radius which is easily seen to diverge when $r \rightarrow 0$. 

\begin{figure}[t]
	\centering
    \includegraphics[scale=0.27]{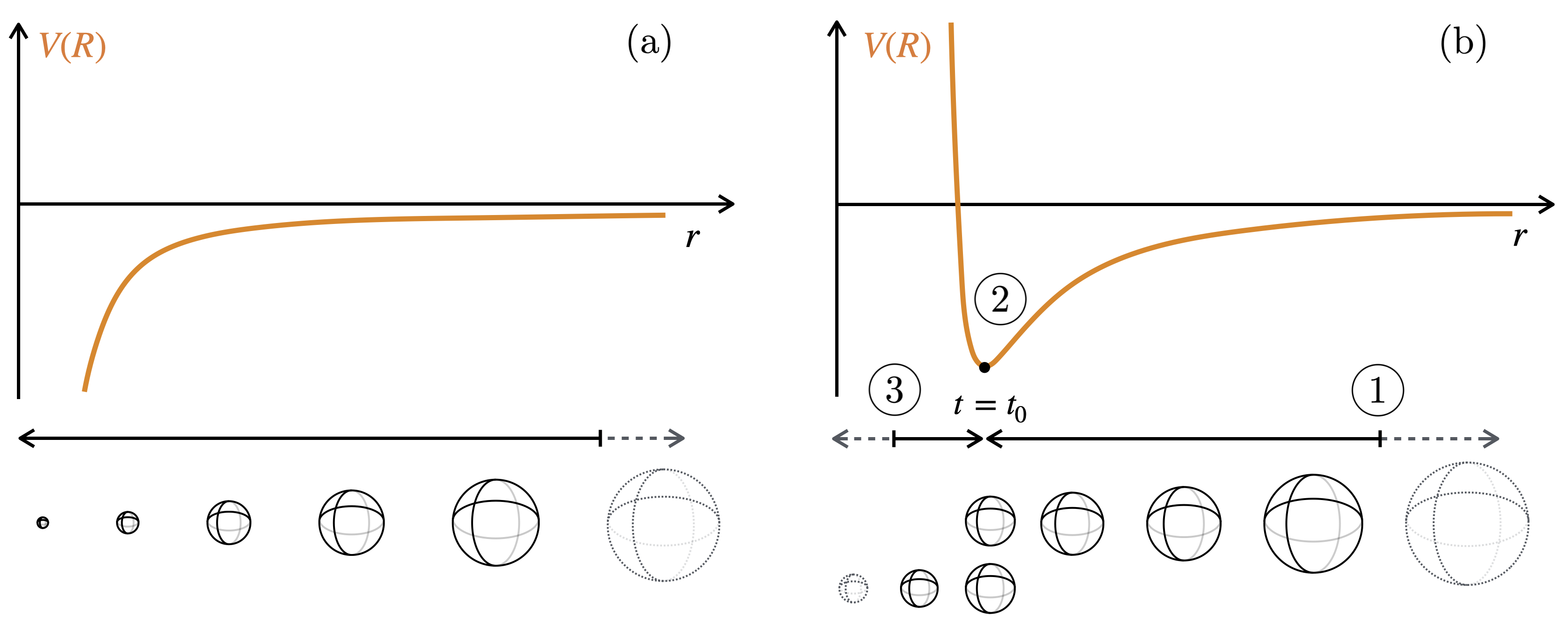}
	\caption{Pictorial summary for generalised Ricci flow and associated potential for $S^3$ with and without $H$-flux, in panel (a) and (b), respectively.}\label{fig:S3_flow_and_potential}
\end{figure}

Here we connect this discussion of the three sphere with flux to the geometric flow picture presented in the introduction. This example was already briefly outlined in section \ref{sec:gRicci_simple_examples} but for convenience we state once more the flow equation for this background which reads
\begin{align}
	\frac{\partial}{\partial t}r(t) = -\frac{2}{r(t)}+\frac{2h^2}{r(t)^4}\,,
\end{align}
where $r$ is the three-sphere radius and $h$ is the three-form flux number and the explicit parametrisation reads
\begin{align}
    \mathrm{d}s^2 = \frac{r^2}{4}\left(\mathrm{d}\eta^2 + \mathrm{d}\xi_1^2 + 4\mathrm{d}\xi_2^2 -4 \cos(\eta)\mathrm{d}\xi_1 \mathrm{d}\xi_2\right)\,,\qquad H = -\frac{h}{2}\sin(\eta)\mathrm{d}\eta \wedge \mathrm{d}\xi_1 \wedge \mathrm{d}\xi_2\,,
\end{align}
with angular coordinates $\eta \in [0,\pi], \xi_1 \in [0,2\pi), \xi_2\in[0,2\pi)$. 
Note that for both radius $r$ smaller than $\sqrt{h}$ as well as $r$ bigger than $\sqrt{h}$ the flow is towards the fixed point at $r=\sqrt{h}$. This can be understood very nicely also in terms of the potential generated by this background (cf. also figure \ref{fig:S3_flow_and_potential}). We simply flow in the potential towards the critical point at $r=\sqrt{h}$.

We can also look at the asymptotic behaviour of this PDE for large and small $r$. In these cases we can neglect the latter or the first term respectively. We get
\begin{align}\label{eq:flow_S3H_rsols}
    \begin{cases}
        r\gg1\,: & \frac{\partial}{\partial t}r(t) \sim -\frac{2}{r(t)} \quad\; \Rightarrow \quad r(t) \sim \sqrt{r_0^2-t},\\
        r\ll 1\,: & \frac{\partial}{\partial t}r(t) \sim +\frac{2h^2}{r(t)^5} \quad \Rightarrow \quad r(t) \sim \sqrt[6]{r_0^6+6 h^2 t}
    \end{cases}\,, \quad  r(0)=r_0\,.
\end{align}
Therefore we see that for large $r$ the flow asymptotes to the one of the sphere without $H$-flux, while for small the flow now diverges with a positive sign. We see from this asymptotic expressions that in principal a flow with initial condition $r(t=0)=r_0$ can admit an extension backwards in time. Solving the system numerically confirms this behaviour for negative $t$. See also figure \ref{fig:S3_H_with_f_backwards}. For $r\gg 1$ this can be done arbitrarily long making the solution eternal, i.e defined for all $t \in (-\infty,\infty)$. For $r\ll 1$ there is a finite time $t^\ast=r_0^6/(6h^2)$ at which radius becomes zero and the potential diverges. Therefore, the solution is ``only'' immortal, with maximal interval of existence $t \in (t^\ast, \infty)$.

We can now turn to determine the distance $\Delta_L$ we defined in section \ref{sec:def_distance} for the various points of interest. Note that, in principle, we calculated the flow in the gauge fixed from, where the scalar $f$ is decoupled from the other flow equations. This is not the gauge in which the unit volume constraint is fulfilled in general so we should be careful when evaluating the formula for $\Delta_L$.\footnote{Of course at the end it should not matter which gauge choice we use, since $\mathcal{F}$ is a gauge invariant quantity. One merely needs to be careful in keeping track of the volume element in case the constant volume constraint is not manifest in some gauge.} However since all the time dependence of $g_{\mu\nu}$ is in the radius and so does not depend on the coordinates,  the scalar field $f$ will also not depend on the coordinates. In these cases, it does not matter which gauge we choose and the constraint will be fulfilled in both flows. 

Recall that the formula for the distance  $\Delta_L$ defined in section \ref{sec:univLength} can be written as
\begin{align}
    \Delta_L = \tfrac{1}{\sqrt{2 V_K}} \int_t\mathrm{d} t  \,\left|\int_K \mathrm d V e^{-f} \Bigl(\left|g^{-1}\partial_t g\right|^2+\left|g^{-1}\partial_tB\right|^2\Bigr) \right|^{1/2}\,.
\end{align}
Since the only time dependence of $g_{\mu\nu}$ is through $r(t)$ we can easily obtain $\left|g^{-1}\partial_t g\right|^2=12 \left(\partial_t r/r\right)^2$  and therefore
\begin{align}\label{eq:length_S3H}
    \Delta_L =  \sqrt{\frac{12}{2}}  \int_{t_i}^{t_f} \left| \left(\frac{\partial_t r}{r}\right)^2\right|^{1/2} \mathrm{d}t=\sqrt{6} \int_{t_i}^{t_f}  \frac{\partial_t r}{r} \mathrm{d}t = \sqrt{6} \log\left(\frac{r(t_f)}{r(t_i)} \right)\,.
\end{align}
We can now evaluate the distance to the three distinguished points in the flow, the fixed point at $r=\sqrt{h}$, the fixed point at $r\to \infty$ and the singularity at $r\to 0$ where the last two can only be reached backwards in time. This yields 
\begin{align}\label{eq:length_S3H_vals}
\begin{cases}
 &\Delta_L(r_0,\sqrt{h}) = \mathrm{const.}\,, \quad \;\,\text{for  $r_0>0$.}\\
 &\;\, \Delta_L(r_0,\infty) = \infty\,, \quad \qquad \text{for  $r_0\geq\sqrt{h}$.}\\
 &\; \;\,\,\Delta_L(r_0,0) = \infty\,, \quad \qquad \text{for $r_0\leq\sqrt{h}$.}  
\end{cases}
\end{align}
We have collected all necessary ingredients in order to finally interpret the situation with respect to the SDC or RFC, respectively. The generalised flow on the three-sphere with three-form flux possesses three distinct regions, see figure \ref{fig:S3_flow_and_potential}, panel (b). In fact, this example, although relatively simple, features already the essential scenarios one can encounter following a geometric flow. Let us describe them one-by-one, whilst juxtaposing with the conclusions arising from studying the scalar potentials as carried out in \cite{Demulder:2023vlo}.

\begin{enumerate}[label=\protect\circled{\arabic*}]
\item Limit $r \to \infty:$ Starting from $r_0 >\sqrt{h}$ and taking the limit $t \to -\infty$ the three-sphere decompactifies as its radius becomes arbitrarily large. Since in this limit, from eq. \eqref{eq:flow_S3H_rsols}, we have that $\lim_{t\rightarrow-\infty}\partial_t r(t)=0$, this is in fact a fixed point of the flow. In addition this point is at infinite distance as measured by the generalised Ricci flow, see eq. \eqref{eq:length_S3H_vals}. We thus obtain a genuine infinite distance point, where the tower of massless states is realised by the associated KK-modes. 
\item At the minimum $r=\sqrt{k}$ of the potential, the flow stabilises the three-sphere at a fixed radius determined by the three-form flux strength. This corresponds to the conformal string vacuum. This is a fixed point of the flow, $\lim_{t_0\rightarrow \infty}\partial_t r(t)=0$. However, the associated distance in eq. \eqref{eq:length_S3H_vals} is finite. This is consistent with our expectations since we have a perfectly well defined completion to Quantum Gravity in terms of string theory and we expect no modes becoming light at this point. Note that this conclusion applies both to the forward and backward flows.
\item Limit $r \to 0$: Now starting from some $r_0 <\sqrt{h}$ and extending backwards the flow encounters a singularity, i.e.  $\lim_{t\rightarrow -\infty}\partial_t r(t)=\infty$, rather than a fixed point. The picture that emerges fits very nicely with that proposed in \cite{Demulder:2023vlo}. 
Compactifying on this configuration one obtains a scalar field space with a non-trivial scalar potential. The scalar potential sources by flux and curvature terms vanishes for large $r$ while diverging for $r \to 0$. The point $r=0$ is not a fixed point and therefore no tower of light states is predicted by the Ricci Flow Conjecture. 
Indeed it was explained in \cite{Demulder:2023vlo} that the divergent potential should be viewed as a signal of a pathological infinite distance limit for a theory that is anyway in the landscape. Hence the potential ``forbids'' us to even visit this limit and therefore no prediction of a tower of light states is made by the SDC in this case. These would usually be given by string winding modes. However the sphere has $\pi_1(S^3)=0$ and there are no non-trivial winding modes that can serve as the tower of light states. The situation we encounter in the present discussion is the manifestation of this very principle, now seen through the lens of (generalised) Ricci flow. 
\end{enumerate}

This example motivates the following reformulation of the Ricci flow Distance Conjecture, taking into account now the effect of a non-trivial flux\footnote{We  discuss RR-fluxes, which require defining an extension of the Ricci flow in section \ref{sec:gen2_Ricci_flow}.}. We propose that by following the geometric flow a tower of states is encountered whenever we approach a fixed point that also lies at infinite distance.

\vspace{-2pt}
\paragraph{Modified Ricci Flow Conjecture.} Consider QG on a family of backgrounds determined by the triple $(g_{ij}(t),B_{ij}(t),f(t))$ satisfying the \textit{generalised} Ricci flow. There exists an infinite tower of states becoming massless along the trajectory of the flow towards a fixed point which \textit{in addition} lies at infinite distance. The distance between two backgrounds at time $t_i$ and $t_f$ is measured in terms of the generalised Perelman entropy functional $\mathcal{F}(g,B,f)$ by $\Delta_L$.\\

Note that infinite distance points with massless towers of states only occur when approaching a fixed point that additionally is associated with a divergent distance by eq. \eqref{eq:length_S3H}, and vice-versa. Indeed, we saw above that these two things do not necessarily go hand in hand. In case 1 above, the distance is infinite. Naively one might indeed expect a tower of light states also in this limit. In the example at hand there are however no modes that can become light in the limit of small $r$.  However the crucial difference is that although the point lies at infinite distance, it is \textit{not} a fixed point of the flow but a singularity.

\paragraph{Remark.} 
\begin{itemize}
    \item The Ricci flow equation is weakly parabolic but can be transformed into a strictly parabolic system using the DeTurck trick -- a method we will revisit when incorporating RR-fluxes in section \ref{sec:gen2_Ricci_flow}. The property of parabolicity ensures that the forward Ricci flow is well-posed, and warrants the existence and uniqueness of short-time solutions. The backward evolution of a parabolic system on the other hand is generally ill-posed. Small perturbations in the metric can be uncontrollably , leading to non-uniqueness and instability in the solutions. Nonetheless, under certain conditions, the backward flow can be well-defined.  These cases are less studied, but \cite{kotschwar2009backwardsuniquenessricciflow} proved for example the uniqueness of solutions to the Ricci flow provided certain boundedness conditions on the curvature. Even fewer results are known for generalisations of the Ricci flow. In the above, and for what is to come, we take a pragmatic stance and assume the backward existence.
\end{itemize}

\subsection{The SDC in the T-dual}
Standard Ricci flow is not invariant under topological T-duality \cite{Garcia-Fernandez:2016ofz,Severa:2018pag}, see also appendix \ref{app:GG_and_T-duality_form}.
A given solution of Ricci flow will in general be mapped under T-duality to a solution of the \textit{generalised} Ricci flow. This will for example happen when T-duality is applied to a non-trivially fibered circle, leading to a dual background with $H$-flux. As was shown in \cite{Garcia-Fernandez:2016ofz}, generalised Ricci flow in contrast is closed under T-duality. Solutions are mapped to other solutions under duality transformation. This implies in particular that for the Ricci Flow Conjecture to be consistent within String Theory, one is forced to extended its formulation to the \textit{generalised} Ricci flow. In this section, we check the consistency of applying the modified Ricci Flow Conjecture to the T-dual of the three-sphere with $H$-flux. The T-dual of the three sphere with $h$ units of $H$-flux follows from the standard Buscher rules and results in the background given by
\begin{gather}\label{eq:metricS2S1}
\begin{aligned}
 \mathrm{d}s^2&=\frac{r^2}{4}\left(\mathrm{d}\eta^2 + \sin(\eta)^2 \mathrm{d}\xi_1^2 \right) + \left(\cos(\eta)\frac{h}{2 r}\mathrm{d}\xi_1 - \frac{1}{r}\mathrm{d}\xi_2 \right)^2\,,\\
     H &= -\frac{1}{2} \sin(\eta) \mathrm{d}\eta \wedge \mathrm{d}\xi_1 \wedge \mathrm{d}\xi_2\,.
\end{aligned}
\end{gather}
The geometry is  locally given by a product of a sphere $S^2$ with radius proportional to $r$ with circle is proportional to $r^{-1}$. Computing the associated flow equation for this background we get 
\begin{align}\label{eq:flows1s2}
	\frac{\partial}{\partial t}r(t) = -\frac{2}{r(t)}+\frac{2h^2}{r(t)^4}\,,\qquad \frac{\partial}{\partial t}h(t) =0\,,
\end{align}
which, as expected, is precisely the flow equation of the system before T-dualising. However, what changes is the geometric interpretation of this background. For generic $h$, the metric above defines a circle fibration over $S^2$, i.e. $S^{1} \hookrightarrow E \to S^2$ . Whenever $h$ is integer (which for us it is because it arises as the T-dual of the integer $H$-flux number), the circle bundle has first Chern class given by $h$, and is known as a Lens space $L(h,1)$. We thus get the same flow equations as before, and in scalar field space the same potential \cite{Demulder:2023vlo}, but a geometrically very different interpretation of the flow and its associated infinite distance points. Referring again to figure \ref{fig:S3_flow_and_potential}:

\begin{enumerate}[label=\protect\circled{\arabic*}]
\item Consider first approaching $r(t) \to \infty$, which can be attained by starting at some $r_0 > \sqrt{h}$ and flowing backwards towards $t\rightarrow-\infty$. Recall that for the flow on the three-sphere with $H$-flux this corresponded to the decompactification. Here though, only the two-sphere factor decompactifies, while the circle fibre vanishes. This is an infinite distance limit and a fixed point of the flow. The relevant tower of states for the SDC comes from the KK modes on the decompactifying two-sphere. One might wonder that contrary to the $S^3$, the circle factor could allow for non-trivial winding and therefore there should be winding modes that become light as the circle shrinks. However, whenever $h$ is non-zero, there is only a finite number of winding states $\mathbb{Z}_h$ and therefore these states do not provide  an additional tower for the SDC. This is actually consistent with the exchange of towers of states under T-duality as we explain in detail in \cite{Demulder:2023vlo}.

\item At the end of the flow given in \eqref{eq:flows1s2}, i.e. taking $t \to \infty$, the radius $r(t)$ stabilises to $\to \sqrt{h}$. This leads to a fixed point consisting of a local product of (up to overall normalisations) a two-sphere with radius $L_{S^2} =\sqrt{h}$ and a circle of radius $L_{S^1}=1/\sqrt{h}$. Note that when $h=1$, the geometry is self-dual under topological T-duality and the fixed point of the flow reduces to the standard round sphere $S^3$ with radius $r=\sqrt{h}=1$, i.e. the conformal point. Although a fixed point, the end of the flow leads to a finite length and thus finite distance for the Swampland Distance Conjecture.

\item On the other hand, in the limit $r(t) \to 0$ the two-sphere vanishes while the circle fibre decompactifies. This is again an infinite distance limit but not a fixed point. Indeed on the $S^2$ factor there are no winding modes that would become light in this limit. On the circle fibre one might again expect an infinite tower of KK modes becoming light as the circle decompactifies. However the presence of non-trivial $H$-flux spoils the conservation of this tower and therefore there is indeed no tower of states for the SDC \cite{Demulder:2023vlo}.
\end{enumerate}

Lastly, let us briefly comment on the transformation of $f$ under T-duality. In the action, $f$ appears as a generalised volume element $\sqrt{g}\,e^{-f}$ and the flow equation for $f$ is obtained by requiring this quantity to be constant along the flow. Denoting the T-dual metric by $\tilde{g}$ we can ask what is the flow equation for a new quantity $\tilde{f}$ such that $\sqrt{\tilde{g}}\,e^{-\tilde{f}}$ is again constant. This is of course nothing else than the original flow equation evaluated in the dual tilde quantities. However, it is a well known fact that the generalised volume element is invariant under T-duality, provided the ``dilaton'' is shifted according to 
\begin{align}
    \tilde{f} = f-\log(\det(g))\,.
\end{align}
From this one can easily derive the new flow equation for $f$ and check explicitly that it gives the right flow in order to keep the generalised volume constant for the T-dual geometry.\footnote{We note that similarly the flow equation for the modified flows can be obtained by simply evaluating the equations in the dual frame or, in case of a gradient flow, by using the T-duality invariance of the volume element under the appropriate shift as done above}

\section{Including RR-fluxes in the geometric flow}\label{sec:gen2_Ricci_flow}

In the last sections we reviewed the generalised Ricci flow and its implication for the Swampland Program as an extension of ``standard'' Ricci flow originally defined by Hamilton. However, considering the full string effective action of type II SUGRA, the inclusion of RR-fluxes is paramount. One could then ask whether the flow can be further generalised to also include the Ramond-Ramond sector, hence including the RR $p$-form fluxes\footnote{See \cite{Kopfer2024} for a closely related flow which, however, cannot be used to study type IIA/B solution as the dilaton, or rather scalar field, is not incorporated in an adequate way into the flow equations.} $F^p$.

In the following we explain this possibility, defining a system of flow equations including also contributions from the RR sector. Closely following the known proof for the NSNS sector, we show that the resulting system is indeed well-defined and investigate the implications for the Distance Conjecture and the Swampland Program.

Defining such a flow is, a priori, highly non-trivial. As noted in the introduction, geometric flows are delicate to construct. A key challenge is that most ``flows'' are not short-lived. Mathematically, this requires the system of partial differential equations to be strictly parabolic, ensuring no singularity arises for arbitrarily small times. In the context of the Distance Conjecture, this condition is crucial for obtaining paths of non-vanishing distance. For the Ricci and generalised Ricci flows, additional properties come into play: both flows are gradient with respect to a functional -- the (generalised) Perelman's functional -- and are strictly monotonous along the flow. While these properties are not strictly necessary for a geometric flow to be well-defined, the existence of a functional is essential for the defined distance to remain meaningful.

A further property, which serves as our guiding principle for including RR-fluxes, is the connection between the flow equations of the (generalised) Ricci flow and the $\beta$-functions or equations of motion. Using this relation as a starting point, we construct a flow that is at least short-lived and for which the Einstein-Hilbert action provides a monotonous functional. Although this action is neither strictly monotonous nor gradient for the flow, these properties are not necessary to define a distance as introduced in section \ref{sec:univLength}.

\subsection{A flow for type II supergravity}\label{sec:flow_typeII}

The flow equations of generalised Ricci flow are nothing else than the equations of motions of NSNS sector of supergravity. We defined a further generalisation of this flow, by including $p$-form RR fluxes such that the resulting flow equations are exactly the type II SUGRA equations of motion. Explicitly, we define for (massive) type IIA theories
\begin{gather}
\begin{aligned}\label{eq:flowRR1}
    \frac{\partial}{\partial t} g_{ij} &=-2\beta^g_{ij}= -2R_{ij}-4\nabla_i \nabla_j \phi + \frac{1}{2}H^2_{ij} \\
    &\qquad\qquad\qquad +2 e^{2\phi}\left[\frac{1}{2}(F_2^2)_{ij} + \frac{1}{12} (F_4^2)_{ij} - \frac{1}{4} g_{ij}\left( \frac{1}{2} F_2^2 +\frac{1}{24} F_4^2 +m^2 \right)\right]\,,\\
        \frac{\partial}{\partial t} B_{ij} &=-2\beta^B_{ij}=\nabla^kH_{kij}- 2H_{kij}\nabla^k\phi+\star( F_2\wedge \star F_4 -\frac{1}{2} F_4\wedge F_4-m \star F_2)_{ij}\,,\\
             \frac{\partial}{\partial t} \phi &=\begin{cases}
            -2\beta^\phi_1=4\Delta \phi -  4(\nabla \phi)^2 +R -\frac{1}{12}H^2\,,\\
            -2\beta^\phi_2=\Delta \phi -  2(\nabla \phi)^2 +\frac{1}{12}H^2+\frac{1}{2}e^{2\phi}\left((1-\frac{n}{4})|F_2|^2+\frac{1}{6}(1-\frac{n}{8})|F_4|^2+\frac{n}{2}m^2\right)\,.
            \end{cases}
\end{aligned}
\end{gather}
The equations for $g,B$ and $\phi$ are supplemented by flow equations for the form potentials
\begin{gather}
\begin{aligned}\label{eq:flowRR2}
             \frac{\partial}{\partial t}C_{i}&=-2\beta^{C_2}_{i}=\left(\nabla^k F_{ki}+\star(H\wedge \star F_4)_{i}\right)\,,\\
                         \frac{\partial}{\partial t}C_{ijm}&=-2\beta^{C_3}_{ijm}=\left(\nabla^k F_{kijm}+\star(H\wedge  F_4)_{ijm}\right)\,.
\end{aligned}
\end{gather}
Similar equations can be defined for type IIB theories by use of the appropriate equation of motions. The choice of the factors for the RR flow equations is in principal arbitrary, although the above choice is convenient in order to obtain flow equations for the fluxes that are similar to the generalised Ricci flow of section \ref{sec:generalised_Ricci}. We use the symbol $\beta^\bullet$ which, however, is purely notational in order to highlight the use of the associated equations of motion like for the NSNS (generalised) Ricci flow. In particular they are not RG flow $\beta$-functions associated to a NLSM description.

Note that we wrote two expressions for the flow of the dilaton. This is due to the fact that there is an ambiguity in defining this particular flow equation. While the first choice $\beta^\phi_1$ corresponds to the equation of motion one gets directly from varying the action, the second choice $\beta^\phi_2$ is obtained by substituting the traced equation $\beta^g$. On shell case these are equivalent, and one often chooses the second form where the Ricci scalar is removed by use of the traced $\beta^g$. The following proof goes through with both choices of flow equations and therefore we simply write $\beta^\phi$.

\subsubsection*{Short time existence and uniqueness}
A priori the flow \eqref{eq:flowRR1}, \eqref{eq:flowRR2} need not be well behaved, i.e. might not allow for even short-time solutions. A direct way to warrant for the existence of solutions of the system of partial differential equations describing the flow is strictly parabolicity\footnote{A system of PDEs $\partial_t u = L u$ is said to be strictly parabolic if the principal symbol of the differential operator $L$ is positive definite. Loosely speaking the principal symbol of $L$ is captured by the highest-order derivatives. For a non-linear $L$ (like $R_{ij}$ viewed as an operator acting on $g_{ij}$) parabolicity is established by working with the linearization of $L$.}. In fact the Ricci flow itself, is not strictly parabolic. In particular,  this non-linear differential equation is degenerate due to the diffeomorphism invariance of the Ricci tensor. In other words, the Ricci flow is not strictly parabolic because the metric can be altered by diffeomorphisms. The process of fixing the gauge, thereby resolves this degeneracy and putting the Ricci flow into a manifestly strictly parabolic form is called the ``DeTurck trick''. Specifically, the Ricci flow equation is modified by adding a term that corresponds to the Lie derivative of the metric along a chosen vector field  $X$
\begin{align}
    \frac{\partial  }{\partial t}g_{ij} = -2 R_{ij} + \mathcal{L}_X g_{ij}\,,
\end{align}
where  $\mathcal{L}_X g_{ij} $ is the Lie derivative of  $g_{ij}$ along  $X$. The vector field $X$  is chosen to depend on the metric $g_{ij}$ and the fixed initial metric  $({g}_0)_{ij}$, as follows
\begin{align}\label{eq:vector_DeTurck}
    X^k = g^{ij} \left( \Gamma_{ij}^k - (\Gamma_0)_{ij}^k \right)\,,
\end{align}
where  $\Gamma_{ij}^k$  and  $(\Gamma_0)_{ij}^k$  are the Christoffel symbols of  $g_{ij}$ and $({g}_0)_{ij}$, respectively. The resulting gauge-fixed Ricci flow takes on the form
\begin{align}
    \frac{\partial }{\partial t}g_{ij}= -2 R_{ij} + \nabla_i X_j + \nabla_j X_i,.
\end{align}
Being a strictly parabolic equation, existence and uniqueness of solutions are then guaranteed for short times via the Nash-Moser implicit function theorem or similar in standard parabolic theory \cite{hamilton1982inverse}. After solving these modified equation, the diffeomorphisms are inverted to recover a solution to the original Ricci flow equation.

We show in this section that the equations in eqs. \eqref{eq:flowRR1} and \eqref{eq:flowRR2}, inspired by type IIA/IIB solutions, when performing a similar DeTurck trick, are in fact parabolic and can apply the existence and uniqueness theorems for parabolic PDEs.

As mentioned in the introduction, (generalised) Ricci flow by itself is not parabolic, due to diffeomorphism invariance. We follow the strategy outlined in \cite{garcia2021generalized}. As outlined above, one first need to identify a vector field to gauge fixing the ambiguity coming from the diffeomorphism invariance. To that effect, one can actually take precisely the same vector field as in eq. \eqref{eq:vector_DeTurck}. 
Furthermore we define differential operators $\mathcal{O}_i(g,B,\phi,F^p)$ consisting of the right-hand side of the flow equations corrected by the Lie derivative of this vector field \footnote{The flow equations of the fluxes $F^p$ are defined in terms of their equations of motion $\beta^{F^p}$ which can be obtained from $\beta^{C_{p-1}}$ by derivation $\mathrm{d}(\cdot)$.} as well as $\nabla \phi$ in order to obtain the analogue of the gauge fixed version of generalised Ricci flow
\begin{gather}
\begin{aligned}
\mathcal{O}_g(g,H,F_p,\phi)&=-2\beta^g + \mathcal{L}_X g-2 \mathcal{L}_{\nabla \phi} g\equiv -2\Tilde{\beta}^g + \mathcal{L}_X g\,,\\
    \mathcal{O}_H(g,H,F_p,\phi)&=-2\beta^H + \mathcal{L}_X H-2 \mathcal{L}_{\nabla \phi} H\equiv -2\Tilde{\beta}^H + \mathcal{L}_X H\,,\\
    \mathcal{O}_{F_p}(g,H,F_p,\phi)&=-2\beta^{F_p} + \mathcal{L}_X F_p-2 \mathcal{L}_{\nabla \phi} F_p\equiv -2 \Tilde{\beta}^{F_p} + \mathcal{L}_X F_p\,,\\
    \mathcal{O}_{\phi}(g,H,F_p,\phi)&=-2\beta^\phi + \mathcal{L}_X \phi -2 \mathcal{L}_{\nabla \phi} \phi=-2\tilde{\beta}^\phi + \mathcal{L}_X \phi\,.
\end{aligned}
\end{gather}
In the last line we have defined $\tilde{\beta}^{g/B}=\beta^{g/B} + \mathcal{L}_{\nabla \phi}(g/B)$, $\tilde{\beta}^{F_p}=\beta^{F_p} + \mathcal{L}_{\nabla \phi}(F^p)$ and $\tilde{\beta^\phi}=\beta^\phi-\mathcal{L}_{\nabla\phi}$.
Note that the Lie derivative with respect to $\nabla \phi$ would usually decouple the dilaton from the rest of the equations, but due to the RR-fluxes this is no longer possible. This is particular to the new flow we are defining and we will return to this fact shortly.

For the next step we need to compute the linearisation of the operators $\mathcal{O}_i$. This is done by formally expanding the metric $g_{ij} = (g_0)_{ij} + h_{ij}$ as well as the other quantities in the DeTruck-modified flow and expand up to first order terms. The resulting PDE is then second order.  To do so, we introduce the following notation/convention. We write $g_s,H_s,F^{p}_s,\phi_s$ in order to explicitly stress the dependence on the linearisation parameter $s$ and where in this section we write $F^p\equiv F_p$. Furthermore defined the shorthand
\begin{align}\label{eq:shorthand}
    \frac{\mathrm{d}}{\mathrm{d}s}\biggr|_{s=0} g_s=\hat{g}\,,\quad \frac{\mathrm{d}}{\mathrm{d}s}\biggr|_{s=0} H_s=\hat{H}\,,\quad \frac{\mathrm{d}}{\mathrm{d}s}\biggr|_{s=0} F^{p}_s=\hat{F}^{p}\,,\quad \frac{\mathrm{d}}{\mathrm{d}s}\biggr|_{s=0} \phi_s=\hat{\phi}\,.
\end{align}
Then it is immediate from generalised Ricci flow on the NSNS sector, that \cite{garcia2021generalized}\footnote{In the following $\Delta$ always denotes the ``rough" connection Laplacian $\Delta=g^{ij}\nabla_i \nabla_j$ while $\Delta_{dR}=-(\mathrm{d}\mathrm{d}^\ast+\mathrm{d}^\ast \mathrm{d})$ denotes the Laplace-de Rham Laplacian. Furthermore we write $\dots$ for all the quantities that we take independent of $t$ for each linearisation, i.e. $\mathcal{O}_g(g_s,\dots)\equiv \mathcal{O}_g(g_s,H,\phi,F^{i_1},\dots F^{i_n})$.}
\begin{gather}
\begin{aligned}
    \frac{\mathrm{d}}{\mathrm{d}s}\biggr|_{s=0} \mathcal{O}_g(g_s,\dots) &= \Delta \hat{g} + \text{l.o.t.}\,,\qquad \frac{\mathrm{d}}{\mathrm{d}s}\biggr|_{s=0} \mathcal{O}_g(H_s,\dots) &= \text{l.o.t.}\,,
\end{aligned}
\end{gather}
and realising that $F^{p}$ (and $\phi$) appear only at zeroth order in derivatives, up to lower order terms, it is clear that
\begin{gather}
\begin{aligned}
    \frac{\mathrm{d}}{\mathrm{d}s}\biggr|_{s=0} \mathcal{O}_g(F^{p}_s,\dots) &= \text{l.o.t.}\,,\qquad \frac{\mathrm{d}}{\mathrm{d}s}\biggr|_{s=0} \mathcal{O}_g(\phi_s,\dots) &= \text{l.o.t.}
\end{aligned}
\end{gather}
Proceeding similarly with the other $\mathcal{O}_{T_i}$, the only leading order contributions are
\begin{gather}
\begin{aligned}
    \frac{\mathrm{d}}{\mathrm{d}s}\biggr|_{s=0} \mathcal{O}_H(g_s,\dots) &= \Xi\,, & \frac{\mathrm{d}}{\mathrm{d}s}\biggr|_{s=0} \mathcal{O}_{\phi}(g_s,\dots) &= \Omega\,,\\
        \frac{\mathrm{d}}{\mathrm{d}s}\biggr|_{s=0} \mathcal{O}_H(H_s,\dots) &= \Delta_{dR} \hat{H} + \text{l.o.t.}\,, &  \frac{\mathrm{d}}{\mathrm{d}s}\biggr|_{s=0} \mathcal{O}_{\phi}(\phi_s,\dots) &= \Delta \hat{\phi}\,,\\
            \frac{\mathrm{d}}{\mathrm{d}s}\biggr|_{s=0} \mathcal{O}_{F_p}(g_s,\dots) &= \Lambda\,, & \frac{\mathrm{d}}{\mathrm{d}s}\biggr|_{s=0} \mathcal{O}_{F_p}(\phi_s,\dots) &= \Sigma\,,\\
                 \frac{\mathrm{d}}{\mathrm{d}s}\biggr|_{s=0} \mathcal{O}_{F_p}(F^p_s,\dots) &=  \Delta_{dR} \hat{F}^{p} + \text{l.o.t.}\,,
\end{aligned}
\end{gather}
where $\Xi,\Lambda,\Sigma,\Omega$ are potentially leading order operators that however drops out of the argument. Now employing the Weizenb\"ock identity for an $n$-form $\omega$, which relates $\Delta_{dR}$ to $\Delta$ via
\begin{align}
    \Delta_{dR}\omega = \Delta \omega + \mathrm{curv}(\omega)\,,
\end{align}
where $\mathrm{curv}(\omega)$ is a term depending on specific contractions of the Riemann tensor with $\omega$, which therefore is of lower order in derivatives on $\omega$. Hence in the above equations we get
\begin{gather}
\begin{aligned}
     \frac{\mathrm{d}}{\mathrm{d}s}\biggr|_{s=0} \mathcal{O}_H(H_s,\dots) &= \Delta \hat{H} + \text{l.o.t.}\,,\qquad
     \frac{\mathrm{d}}{\mathrm{d}s}\biggr|_{s=0} \mathcal{O}_{F_p}(F_s^p,\dots) &= \Delta \hat{F}^{p} + \text{l.o.t.}\,.
\end{aligned}
\end{gather}
With this we can write the principal symbol $\sigma$, in a basis defined by \eqref{eq:shorthand}, as
\begin{align}
    \sigma \left( D_{(g,H,\phi,F^{p})} (\mathcal{O}_g,\mathcal{O}_B,\mathcal{O}_\phi,\mathcal{O}_{F^p})\right)\begin{pmatrix}
           \hat{g} \\
           \hat{H} \\
           \hat{\phi} \\
           \hat{F}^{p}
         \end{pmatrix} = \begin{pmatrix}
           \Delta & 0 & 0 & 0 \\
           \Xi & \Delta & 0 & 0 \\
           \Omega & 0 & \Delta & 0 \\
          \Lambda & 0 & \Sigma &  \Delta \\
         \end{pmatrix} \begin{pmatrix}
           \hat{g} \\
           \hat{H} \\
           \hat{\phi} \\
           \hat{F}^{p}
         \end{pmatrix} \,,
\end{align}
where the last entry $\Delta$ should be again read as a block matrix, and similarly, the other $(4,i)$ and $(i,4)$ entries as column or row vectors. From this we see, using Silvester's criterion, that $[\sigma_{ij}]$ is positive definite. We established the short time existence of the gauge-fixed flow.

In a second step we need to undo the gauge modification of the flow  by applying a specific diffeomorphism-gauge transformation,  bringing it to the ``standard" generalized Ricci flow form given in terms of the $\Tilde{\beta}_i$. Again following \cite{garcia2021generalized} we define a one-parameter family of diffeomorphisms $\psi_t$ via
\begin{align}
    \frac{\partial}{\partial t}\psi_t &= - X(g_t,g_0) \circ \psi_t\,,\quad \text{with} \;\, \psi_0 = \mathrm{id}\,,
\end{align}
which inverts the DeTurck gauge fix and define the pulled back fields
\begin{align}
    \tilde{g}_t &= \psi^\ast_t g_t\,,\qquad  \tilde{F}^{p}_t = \psi^\ast_t F^p_t\,, \qquad 
    \tilde{H}_t= \psi^\ast_t H_t\,, \qquad  \tilde{\phi}_t = \psi^\ast_t \phi_t\,.
\end{align}
Following \cite{garcia2021generalized}, a short computation\footnote{We use the identity for time dependent vector fields and tensors $\frac{\mathrm{d}}{\mathrm{d}t}(\psi^\ast_t T_t)=\psi^\ast\left(\mathcal{L}_{X_t}T_t+\frac{\mathrm{d}}{\mathrm{d}t}T_t \right)$ (cf. \cite{lee2012smooth}, Prop. 22.15)} shows 
\begin{align}
    \partial_t \tilde{g}_t 
    &=\psi^\ast\left( -2 \Tilde{\beta}_{g}(g,H,F^p,\phi)+\mathcal{L}_{X(g_t,g_0)}g)\right)+ \psi^{\ast}\mathcal{L}_{-X(g_t,g_0)}g_t\nonumber\\
    &= -2 \Tilde{\beta}_{g}(\Tilde{g},\Tilde{H},\Tilde{F}^{p},\Tilde{\phi})\,,
\end{align}
which trivially carries then over also for pulled back RR-fluxes $\Tilde{F}^p$ and dilaton field $\Tilde{\phi}$.
Therefore we showed that the quantities $\{\Tilde{g},\Tilde{H},\Tilde{F}^{p},\Tilde{\phi}\}$ satisfy geometric flow equations extending the conventional generalised Ricci flow to include RR-fluxes defined by the $\{ \tilde{\beta}^\bullet \}$. This completes the proof of existence. Furthermore by applying the ``inverse'' of $\mathcal{L}_{\nabla \phi}(\cdot)$ this establishes the existence of a family of backgrounds $\{\Bar{g},\Bar{H},\Bar{F}^{p},\Bar{\phi}\}$ satisfying the flow in the gauge defined by the $\{ \beta^\bullet \}$, hence the equations of motion of type IIA\footnote{The proof carries over directly also to type IIB.} SUGRA.

We can translate the flow equations for the form fields back into the ones corresponding to their potentials, by simple integration. This yields for example for $\Tilde{H}=H_0+\mathrm{d}\Tilde{B}$ 
\begin{align}
 2\partial_t \mathrm{d}\Tilde{B}_t&=2\partial_t \Tilde{H}_t=\Delta_\mathrm{dR} \Tilde{H} +\mathrm{d}\ast( \Tilde{F}_2\wedge \star \Tilde{F}_4 + \frac{1}{2}\Tilde{F}_4\wedge \Tilde{F}_4-m \star \tilde{F}_2)\,,
\end{align}
such that we obtain 
\begin{align}
     2\partial_t \Tilde{B}_t=-\mathrm{d}^\ast \Tilde{H} +\ast(\Tilde{F}_2\wedge \star \Tilde{F}_4 + \frac{1}{2}\Tilde{F}_4\wedge \Tilde{F}_4-m \star \tilde{F}_2)\,.
\end{align}
Similarly the equations for the RR-fluxes can be integrated.

\paragraph{Remark.}  
\begin{itemize}
\item Alternatively, one could vary the type II action functional, following the approach taken  for the purely NSNS sector (see appendix \ref{app:modified_flows} for additional details). However the critical observation is that the additional contributions from the RR fields can be repackaged into the flow equation of the metric g. Therefore we again encounter as the problematic contribution the term that reads
    \begin{align}
        S(g,B,f)\left(\frac{1}{2}g^{ij}\frac{\partial g_{ij}}{\partial t}-\frac{\partial f}{\partial t} \right)\,.
    \end{align}
Hence, it appears possible to generate a gradient flow by imposing the  (unaltered) generalized unit volume constraint once again.
Following these steps, the resulting equation would precisely match those given in eq. \eqref{eq:flowRR1}, except for the dilaton $\phi$ (or, equivalently, scalar field $f$). The latter would then inherit a flow equation from the trace of $\beta^g$, in a way such that the above contribution vanishes. This precise flow equation is highly problematic as it contributes negatively to the principal symbol, rendering it indefinite. While performing the proof for the purely NSNS sector once can decouple $f$ from the system of PDEs and therefore show the short-time existence of solutions like above. This decoupling is no longer possible including the RR fluxes. This directly implies that, while the usual short-time existence proof applies automatically to generalized Ricci flow, once extended with RR-fluxes the flow becomes potentially ill-behaved. We will therefore not pursue this possibility here.
\end{itemize}

\subsection{A flow for 11D supergravity}

An arguably even simpler case for which we can attempt to include RR fluxes is within the setting of 11D Supergravity. Indeed, in 11D there is no dilaton implying that the flux terms are no longer weighted differently than the Ricci scalar. This simple difference, compared to the type IIA equations \eqref{eq:flowRR1} and \eqref{eq:flowRR2}, drastically simplifies the flow. The absence of a dilaton however comes as a two-edged sword as there is clearly no way to impose a generalised unit volume constraint. In particular, the flow will generically not be gradient or monotonous, an essential ingredient in defining a notion of length. We will return to what this implies in terms of defining a distance in the spirit of section \ref{sec:univLength} in the following section. 
For now,  recall the action of 11-dimensional supergravity, which reads
\begin{align}
    S = \frac{1}{2 \kappa_{11}^2} \int \mathrm{d}^{11}x\sqrt{|g|}\left(R-\frac{1}{2}|F_4|^2\right)\,.
\end{align}
where $\kappa_{11}$ is the $11$D gravitational coupling, $R$ the  Ricci scalar and $F_4$ the four-form flux.  

Following the same reasoning as in the previous section, we define a flow for the metric and $g$ and four-form flux $F_4$ (or $C_3$) according to the equations of motion of this theory. This leads to the flow equations\footnote{A flow of the same form has in fact already been proposed in \cite{Fei:2018mzf}, see also \cite{Fei:2021pbw}. There however, the \textit{total space} is flowed rather than only the internal space of the 11-dimensional background. Since the metric is then of Lorentzian signature, the short time existence of the flow is not guaranteed in general.}  
\begin{align}
    \frac{\partial}{\partial t} g_{ij} &=-2\beta^g_{ij}= -2\left(R_{ij} - \frac{1}{12}(F_4)^2_{ij} + \frac{1}{144}|F_4|^2 g_{ij}\right)\,,\\
    \frac{\partial}{\partial t} C_{ijm} &=-2\beta^{C_3}_{ijm}=\left(\nabla^k F_{kijm} + \frac{1}{2}\star(F_4 \wedge F_4) _{ijm}\right)\,.
\end{align}
As before the symbol $\beta^{\bullet}$ is purely notational as these are not $\beta$-functions in the usual sense.

The short-time existence directly follows from the discussion in the last section but now without a dilaton contribution.

\subsubsection*{Example: Flows with RR-flux on $S^4$}
A simple example illustrating the flow of the internal space of an 11-dimensional supergravity background is provided by the  four-sphere, supported with $f_4$ units of $F_4$ flux. The resulting flow equations are very similar to the NSNS generalised Ricci flow. We obtain
\begin{align}
    \frac{\partial}{\partial t} r(t) = -\frac{3}{r(t)^2}+\frac{f_4^2}{3 r(t)^7}\,, \qquad \frac{\partial}{\partial t} f_4 = 0\,.
\end{align}
Note that, as was already the case of the generalised Ricci flow in section \ref{sec:gen2_Ricci_flow}, since the flux is proportional to volume forms, does not flow. The 4-form flux does however contribute to the flow of the metric, stabilising the sphere at a finite length related to its field strength. 

\begin{figure}[t]
	\centering
	\includegraphics[scale=0.6]{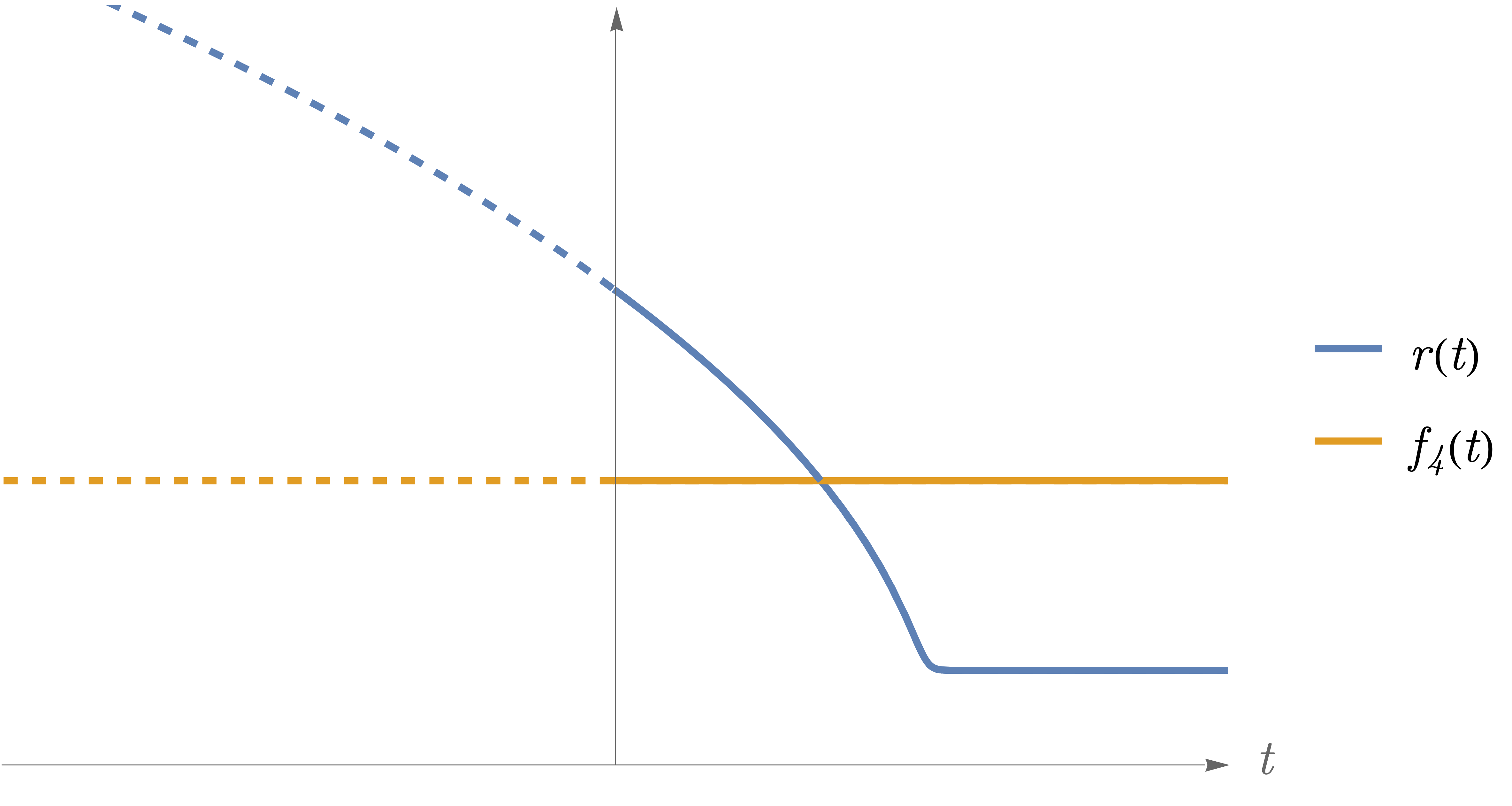}
    \caption{This figure illustrates the flow of the $S^4$ background with $f_4=3$ units of RR flux, serving as the internal part of an $11$D SUGRA background. The plot shows the radius of the sphere, $r(t)$ (blue), stabilising at $r=1$ as $t \to \infty$, and the flux (orange), which remains constant under the flow. The dashed lines depict the extension of the flow backwards in time.}\label{fig:S4_F4_backwards}
\end{figure}

The resulting picture (cf. figure \ref{fig:S4_F4_backwards}) for the SDC closely resembles that of the three-sphere with NSNS flux discussed in section \ref{sec:potentials_and_flow}. There are two infinite distance limits at $r \to \{0,\infty\}$ but only one of them  -- the $r \to \infty$ limit --  is also a fixed point, while $r \to 0 $ is a singularity. Again we can expect KK-modes in the decompactification limit while in the small $r$ limit we are lacking winding modes. However, the main difference to the case studied in section \ref{sec:potentials_and_flow} is, that this geometry solves only the 4-dimensional subset of equations of motion of $11$D SUGRA. In other words, it cannot be trivially completed to a full 11-dimensional solution with a flat external space. It can, however, be completed to a solution of the Freund-Rubin type by the addition of  an external AdS space of the appropriate radius. In general, this larger space will significantly  alter the set of flow equations. Therefore, and in contrast to the NSNS sector 3-sphere with flux, we cannot make definite statements for the Swampland program as this would require treating full product spaces. Hence this example is merely a first step in exploring the possibility of adding higher form fluxes. We plan to come back to this problem in the near future. 

Lastly, we stress again that in $11$D there is no dilaton and therefore no obvious way to implement a generalised constant volume constraint and thus gradient flow. A possible approach\footnote{We would like to thank Chris Blair for a very insightful discussion, in which he pointed out this possibility to us.} here, could be to consider a warping factor in the external metric, in a way such that the problematic trace part of the metric is removed from the flow equations, which in turn upon reducing to 10D then can be seen as a dilaton.

\section{SDC in scalar field space: distance or cost?}\label{sec:distance_vs_cost}

 Recent work pointed out that in the presence of fluxes or equivalently when the moduli space is lifted by a potential, the measure pertinent to the Distance Conjecture may no longer be a bona-fide distance. This was for example the case in \cite{Mohseni:2024njl}, where the defined distance, although passing several non-trivial tests, fails to verify e.g. the triangle inequality. In addition, the requirement for a \textit{geodesic} distance needs to be reconsidered. Indeed, potentials introduce ``external forces'' in moduli space motion which drive the trajectories to non-geodesic paths \cite{Calderon-Infante:2020dhm}. 

\subsection{Properties and non-properties of the distance}

In the original formulation of the SDC, the requirement of geodesicity was tantamount to minimising the path. 
In this section however, we carefully lay out how the length defined in section \ref{sec:univLength} in turn fails and succeeds in being a genuine geodesic distance. Naturally leading us to conclude that, the Ricci Flow Conjecture favours visiting distance according to a minimising cost principle, rather than a distance.

\paragraph{Unidirectionality.} In all these cases considered so far a non-trivial potential in the scalar field space introduces a preferred direction. The similar plight affects the Ricci Flow Conjecture: in general the flow can only go one way. Indeed, the Ricci flow can be seen as a generalised heat or reaction-diffusion equation and a backwards heat equation is generically ill-defined.  In addition, the flow cannot connect arbitrary points in the manifold of possible (generalised) metrics, cf. ``Triangle inequality'' below.

\paragraph{Incompleteness.} In the original Distance Conjecture, the distance is measured along geodesic paths. Since most moduli spaces are described by spaces which are geodesically complete, the ``search'' one can perform in such a way is exhaustive. In contrast this is distinctively no longer true for the Ricci Flow Conjecture.  While any point can serve as an initial condition to the flow, its final point has to be a fixed point or critical points of Perelman's functional. 

 \begin{figure}[t]
    \centering
    \includegraphics[scale=0.21]{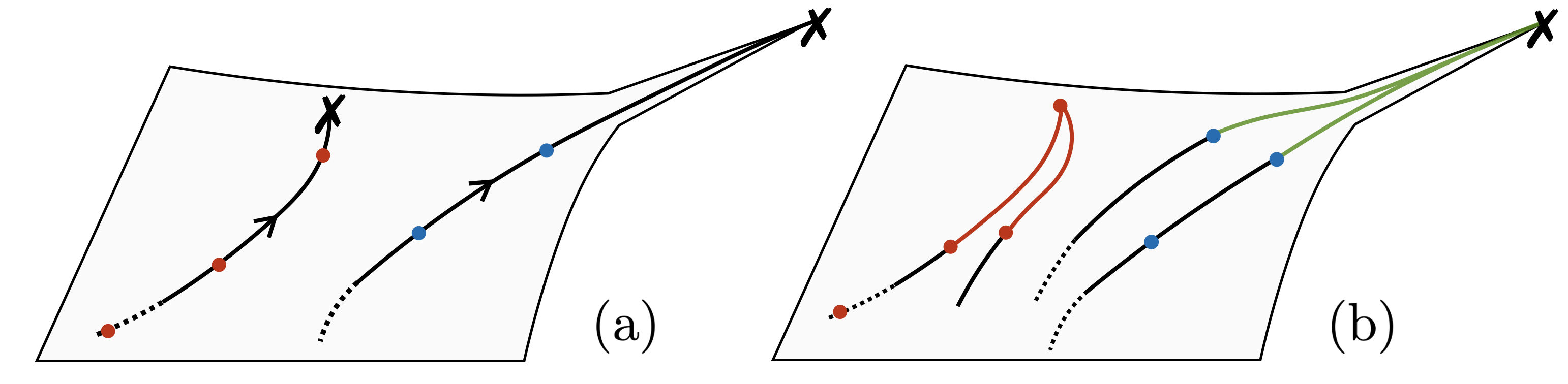}
	\caption{Ricci flow singles out the possible paths one can traverse in field space. Hence in some sense the distance is only defined for points that lie along the same orbit. While distances between red dots or between blue dots can be assigned there is no flow connecting blue points to red points. If one allows to measure distances through fixed points, we can assign a distance to the red dots that lie on different Ricci flow trajectories. This is only possible for fixed points at finite distance. The fixed point of the flow with blue points is at infinite distance and it cannot be used to assign a distance to the blue dots on different curves.} \label{fig:triangle_ineq_reduced}
\end{figure}

\begin{figure}[t]
    \centering
	\includegraphics[scale=0.25]{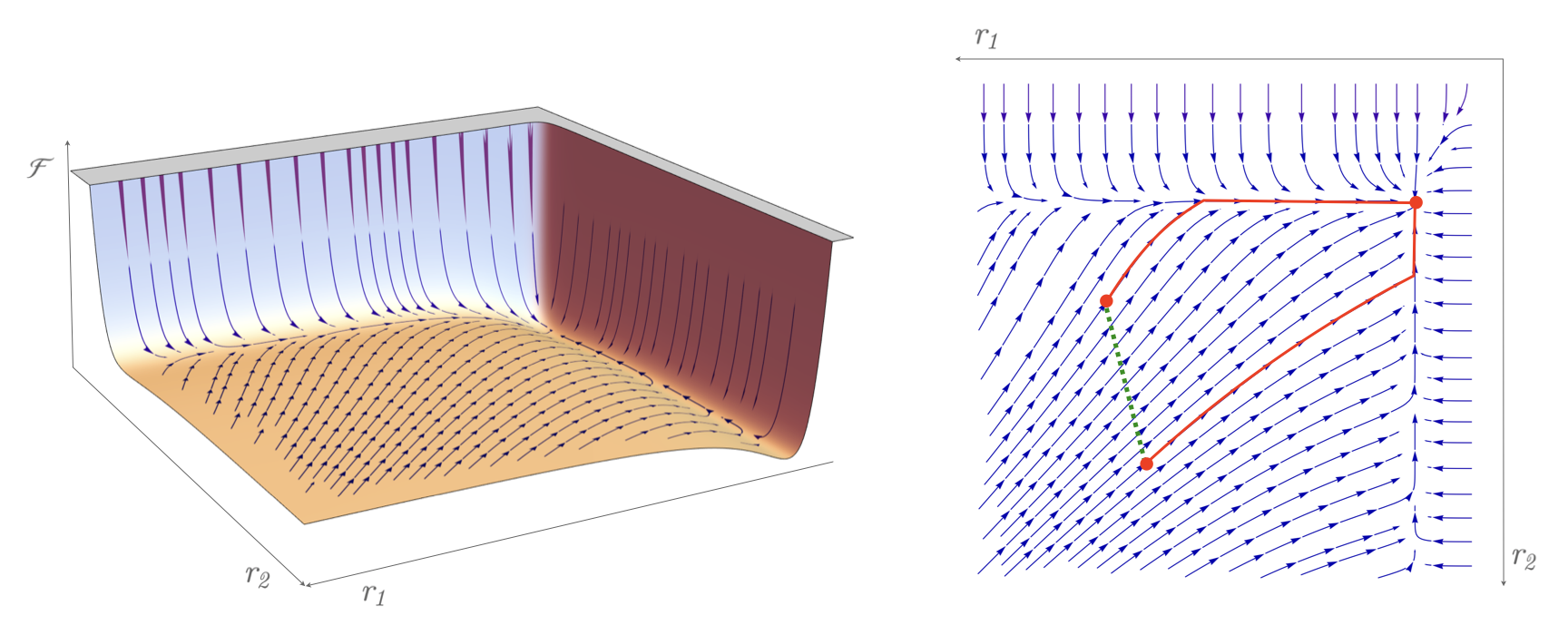}
	\caption{Illustration of the principle explained in the paragraph ``Triangle inequality'' on the example of an internal space of the form $S^3\times S^3$ with each supported by a three-form $H$-flux. The geometry is a direct product. Choosing appropriate initial conditions one can have a flow through all    possible pair of radii ${r_1,r_2}$ of the two spheres. The plot shows the flow for flux numbers ${k_1=1,k_2=3}$. In order to measure the distance between two points using paths selected by Ricci flow one needs to go through the common fixed point at ($r_1=1,r_2=3$). The dotted green line corresponds to the shortest distance in the purely mathematical sense, hence only with respect to the natural metric on the space of metrics.}
    \label{fig_S3_S3_reduced}
\end{figure}

\paragraph{Triangle inequality.}  One of the fundamental properties\footnote{(Strict) positivity is guaranteed whenever the functional is (strictly) monotonous along the flow i.e. $\Delta_L(x,y)>0$ for $x\neq y$ and $\Delta_L(x,x)=0$. Furthermore we have symmetry, i.e. $\Delta_L(x,y)=\Delta_L(y,x)$, where we can think of traversing the path ``backwards in time".} of a proper mathematical notion of distance is the triangle inequality, i.e. $\Delta_L(x,y)\leq \Delta_L(x,z)+\Delta_L(z,y)$. The Ricci flow draws only for specific one-parameter paths in the space of (generalised) Riemannian metrics. While three points along this path (if backwards flow is allowed) can be connected, this is no longer true for randomly selected points in the space of (generalised) Riemannian metrics \ref{fig:triangle_ineq_reduced} a). The only possibility of connecting two points that lie on different Ricci flow curves, is by ``connecting" them at a fixed point, at which two different flow lines merge. Of course here one has to be careful. While a priori there is no problem in doing this when the fixed point is at finite distance, it surely is problematic when trying to do so with a fixed point at infinite distance, cf. \ref{fig:triangle_ineq_reduced} b). However due to their very nature of lying at infinite distance, it appears natural to exclude these fixed points from the reasoning.  As a concrete example of such a situation, see figure \ref{fig_S3_S3_reduced}. Furthermore, we stress that the space of generalised metrics admits a natural metric. Hence, in principle, a distance (in the mathematical sense) can also be defined for points not lying on the same flow orbit. However, it remains unclear how these paths translate to scalar field space with a potential. Indeed, certain paths may not be realised from a minimising or other first-principle criterion. This stands in contrast to the paths singled out by generalised Ricci flow, which have a clear interpretation as being gradient with respect to $\mathcal{S}_{\mathrm{EH}}$.

\paragraph{Strict monotonicity.}  The monotonicity properties of the length or distance measure defined in section \ref{sec:univLength} is directly inherited  from the monotonicity of the flow functional. While the distance is monotonous by definition the property of strict monotonicity depends on the functional. We saw two cases for  which the flow functional itself loses its monotonicity property. First in section \ref{sec:distance_non_gradient}, when weakening the unit volume constraint, the resulting flows where generically not even gradient. The second in section \ref{sec:gen2_Ricci_flow} was encountered when invoking a flow for type IIA/B and 11D supergravity. One could alternatively only demand \textit{weak} monotonicity, in which the functionals together with the the distance $\Delta_L$ remain viable in both cases. Only demanding for a weak, generically progression along the flow may sometimes not go hand in hand with an increase in length.  It would be interesting to understand which precise role the property of \textit{strict} versus \textit{weak} monotonicity plays in the statement of the Ricci flow and Distance Conjectures.

   \paragraph{Geodesicity.} The space of Riemannian metrics admits a set of geodesics with respect to its natural metric, the deWitt metric.\footnote{See \cite{DeBiasio:2023hzo} for a detailed derivation and enlightening discussion in the context of Ricci flow. Let us also remark that moduli spaces generically admit several distance measures, which may not necessarily lead to the same prediction for the SDC, see \cite{Li:2021utg} for a concrete example.} Although the Ricci flow naturally arises as a gradient flow with respect to the same deWitt metric, the path followed by the Ricci flow are in general not geodesic. In other words, the path selected by the flow should be seen as minimising the relevant flow functional, but which may not coincide with a geodesic path, at least with respect to the deWitt metric. This remains of course true for the generalised Ricci flow.

\vspace{10pt}

These observations leave us with the following dilemma. Should one insist that a genuine distance underpins the formulation of the Swampland Distance Conjecture or should one instead favour a cost function instead? In the next section, we contrast these conclusions with other recently proposed approaches.

\subsection{Relation to other notion of distance in scalar field space}\label{sec:rel_dist}

We briefly outline possible ways to connect our notion of distance to existing work on the SDC in the presence of a potential on scalar field spaces. 

When studying the geometric flow in the context of string compactifications, one really has two options. The first, which is the main topic of this paper, is flowing the compact space or total space.  In section \ref{sec:ricci_SDC}, we explained how starting from a $D$-dimensional theory the action splits into an ``external" and ``internal" part and subsequently focused on the internal geometry. The second, alternative option, is to work with the effective action obtained after compactification. Indeed, upon compactifying the internal manifold $K_{n}$ one obtains a lower-dimensional (effective) action 
\begin{align}\label{eq:effective_action}
\mathcal{S}_\mathrm{eff}=\int\mathrm d ^d x\,\sqrt{|\hat{g}|}e^{-2\hat{\phi}}\left(R(\hat{g})-\frac{1}{2}g_{ij}\partial_\mu \varphi^i\partial^\mu \varphi^j+V(\varphi^i)\right)\,,
\end{align}
where $\varphi^i$ are the moduli parametrising the internal space. In the special case of taking $K_{n}$ to be Calabi-Yau, the kinetic term for these fields is controlled by a K\"ahler metric $g_{I\Tilde J}=\partial_I\partial_{\Tilde J}\mathcal K\,,$ where $\mathcal K$ is the K\"ahler potential associated to the Calabi-Yau manifold $K_{n}$. One could now attempt (whenever the scalar potential is zero $V=0$) to flow the moduli space, i.e. the geometry defined by the kinetic term $g_{ij}$ which would give an alternative way to explore genuine moduli spaces. Alternatively one might try to define a flow directly for the system \eqref{eq:effective_action} taking into account also the scalar potential \cite{DeBiasio:2023hzo} (and the external space parametrised by $\hat{g}$.).  While there has already been some work in defining flows for such an action, the resulting flows will in general not be as nicely behaved as flows that evolve the geometry directly. Furthermore, especially when taking the external space into account, one has to properly account for the Lorentzian signature, which may lead to further complications. Since however an action of the form \eqref{eq:effective_action} is the starting point of the considerations in \cite{Mohseni:2024njl} and \cite{Debusschere:2024rmi} it is clear that a direct comparison with the notion of distance we define in this paper and the one in the listed publications is non-trivial.  

However, we stress that in \cite{Kehagias:2019akr} it was observed that when applied to external AdS-spaces, Ricci flow and its associated conjecture successfully recovers the AdS Distance Conjecture, and the associated distance in the presence of a cosmological constant. Varying the action with respect to the Weyl factor, one sees that it is affinely related to the Ricci flow parameter. In particular, the distance from an initial AdS space to the flat fixed point space diverges logarithmically with respect to the Ricci flow parameter $t$ , or linearly with the Weyl rescaling parameter  $\tau$, placing it at an infinite distance
\begin{align}
    \Delta \simeq \log \left(\frac{\Lambda(t_f)}{\Lambda(t_i)} \right)\,,\quad \Lambda(t_i)=\frac{\Lambda_0}{1-2\Lambda_0 t}\,.
\end{align}

We note that, a priori, neither of the two flow scenarios may be entirely satisfactory as it is unlikely that one can probe the whole scalar spaces in either way\footnote{Although, in general we are able to cover the whole moduli space, by picking any point as an initial condition, the possible trajectories will be constrained. A clear example is the K\"ahler-Ricci flow. Any trajectory generated from this flow will preserve the complex structure. In other words we cannot look for infinite distances in the complex structure moduli using this flow.}. In principle one could use other geometric flows, which do not result from the compactification process to probe other trajectories. 

In \cite{Bonnefoy:2019nzv,Cribiori:2022cho,Cribiori:2022nke,Delgado:2022dkz}, the authors used the attractor mechanism to probe the moduli space when the associated space admits  supersymmetric BPS states. In addition, the distance was related to the difference in squared ADM mass of the black hole. Their method is in spirit very similar to the first scenario, where one flows the internal space, here a Calabi-Yau manifold, by the means of a mechanism that gradually changes the internal moduli. In that case the attractor equations. The crucial difference is that the attractor mechanism is, in general, not a geometric flow. One could however hope to find a geometric flow which, although visiting different path in moduli space, shares the same infinite distance points. Provided such a flow exists,  distances can be  measured through its entropy functional.

The second perspective involves working directly at the level of the scalar field space with potential that is the starting point of the considerations in \cite{Mohseni:2024njl,Debusschere:2024rmi}. Relating our approach to such an analysis would require working with a flow of the scalar field space metric $g_{ij}$ sourced by the scalar potential. However, generically this will not lead to a well-defined flow since the equations will generically fail to be (weakly) parabolic.

\section{Conclusions}\label{sec:conclusions}

This work is a natural continuation of \cite{Kehagias:2019akr}, where the Swampland Ricci flow conjecture was originally conceived. Motivated by insisting on consistency under T-duality and the abundance of string solutions with fluxes, we showed how the conjecture extends to flux-supported internal spaces. This perspective on the Distance Conjecture in fact suggests a new way to explore scalar field spaces with potentials. We propose to systematically explore the space of (generalised) Riemannian metrics and RR-fluxes, utilising geometric flows. Indeed, geometric flows are often motivated by the wish to reduce complicated geometries to simpler and canonical ones, for example, Ricci flat or Calabi-Yau manifolds. Coincidentally, these canonical geometries naturally arise in string theory. In particular, using geometric flows, one effectively geometrises the exploration of scalar field space. It bypasses the need to perform the compactification procedure leading to the metric and potential on scalar field, in favour of varying the metric of the internal space and the supporting fluxes through flow equations. Each flow line in the space of metrics draws out a trajectory in the corresponding scalar field space. Indeed, rather than directly measuring distances in scalar field space, we now measure the length of the path taken by the flow on the (abstract) space of generalised metric in terms of the gradient functional governing the flow. In addition this approach provides a natural way to construct a controlled off-shell path that returns to an on-shell configuration exactly at the flow's fixed points.

In this paper, we employed this strategy to explore scalar fields in internal spaces supported by fluxes, as they naturally emerge from string compactifications, therefore mainly focusing on String Theory field space. We first generalised the definition of a distance appearing in the Ricci flow conjecture to geometric flows that are gradient. For the special case where the gradient functional is the generalised Perelman entropy, we show how it is directly related to the deWitt or Candelas-de la Ossa metric \cite{DeWitt:1967yk,Candelas:1990pi} on the space of generalised metrics. In addition, we show how this generalised distance offers a first principle derivation of  logarithmic  distance with respect to the Pereleman functional used in the original formulation of the Ricci Flow Conjecture \cite{Kehagias:2019akr}.

We then extended the conjecture to allow for non-trivial Kalb-Ramond fields by utilising the generalised Ricci flow of \cite{Garcia-Fernandez:2016ofz,streets2017generalized,garcia2021generalized}. This flow already highlights the key distinction from standard Ricci flow. Fluxes generically stabilise the flow, where otherwise the geometry would hit a singularity. We showed how the generalised Ricci flow, and its associated length, successfully probe the infinite distance points found in \cite{Demulder:2023vlo}. In fact, the stabilising property of the flux perfectly mirrors the behaviour of the potential. The divergence in the potential at infinite distance translates to an infinite-distance point along the generalised Ricci flow, which is not a fixed point of the flow.

In order to extend the Ricci Flow Conjecture to account for RR-fluxes, we then consider a flow capturing internal spaces in IIA/B and 11D supergravity. We discuss how, while the flow admits short-time solutions, it loses its gradient property. Although one can in principle insist on defining a length using the non-gradient associated functional, in general it will not be monotonic along the path followed by the flow.   The situation is even more challenging for 11D supergravity, where in absence of a dilaton, no unit-volume constraint can even be imposed to regularise the flow.

\section*{Future perspectives}
This work suggests a more general narrative than the standard approach to moduli spaces or scalar field spaces with potential in the context of the SDC. For flux-supported geometries, or equivalently for a scalar field spaces with a potential, infinite distances can be probed  using a set of geometric flows. The path is singled out by Ricci flow and its corresponding gradient functional. The latter allows for a very natural definition of the distance that agrees with the conventional notion of distance in the space of generalised metrics. Here, we outline a few direct extensions and promising future directions stemming from our results.

\paragraph{Product spaces, DGKT and quantum corrections.} Unlike for the standard Ricci flow \cite{DeBiasio:2020xkv}, the presence of fluxes opens up the possibility of applying generalised Ricci flow to product spaces and therefore full-fledged string vacua. It would be interesting to apply the same ideas presented here to full type II(A) solutions, where the moduli can be fully stabilised and one can discuss aspects like scale separation. In particular, this opens up the possibility to test backgrounds like DGKT via the Ricci Flow Conjecture and their status in the Swampland program \cite{Palti:2024voy,Montero:2024qtz}. With the ability of tackling those examples, come many new and interesting aspects, like the addition of localised or smeared sources, non-trivial warping and also -- as discussed in detail in \cite{Montero:2024qtz} --  the role of quantum corrections. Through the Ricci Flow Conjecture, it might then be possible to access quantum corrections by identifying the flow equations with an RG flow, for which these corrections have been studied using integrability techniques \cite{Hoare:2019ark}.

\paragraph{$N=2$ flux compactifications and generalised Calabi-Yau manifolds.} In \cite{Grimm:2018ohb,Corvilain:2018lgw}, infinite distances in the moduli space of complex structure were identified by their infinite order monodromy matrix. One could wonder if these infinite distance points are realised as the fixed points of an adequate flow. An obvious candidate is the K\"ahler-Ricci flow \cite{cao1986deformation} which naturally varies K\"ahler moduli while keeping complex moduli fixed. 

Geometric flows in addition suggest how this discussion can be in fact be extended to complex moduli and combined variations. Going beyond Calabi-Yau manifolds, the K\"ahler and complex moduli can no longer be disentangled. The resulting background fields are then most naturally described in terms of pure spinors and complex geometry. The generalised Ricci flow in fact admits a particularly elegant description in terms of generalised and complex geometry \cite{garcia2021generalized}. Even more, by combining two different generalised Ricci flows allows one to naturally flow generalised K\"ahler manifolds. Dubbed the generalised K\"ahler-Ricci flow \cite{apostolov2022generalized}, the flow admits an elegant formulation in terms of pure spinors and a functional closely related to Hitchin's functional. Using this flow one can start to probe the scalar field space of $N=2$ flux compactification, hence going beyond Calabi-Yau compactifications. In addition, the known relation between the generalised Hitchin functional and $N=2$ compactifications  \cite{Grana:2005ny,Benmachiche:2006df} as well as between this functional and the attractor mechanism \cite{Pestun:2005ni,Hsu:2006vw}, offer a promising avenue to connect the Ricci Flow Conjecture to the Black Hole Entropy Distance Conjecture \cite{Bonnefoy:2019nzv} and extend it beyond Calabi-Yau compactifications. We plan on reporting on these points soon \cite{followup}.

\paragraph{Existence of dualities and scalar field space.} The recent paper  \cite{Delgado:2024skw} explores the relationship between finiteness of quantum-gravity amplitudes and the existence of dualities, amongst which T-duality, through the compactifiability of moduli spaces. The authors argue that this implies the necessity for semisimple representations within duality groups. These results could be extended to scalar spaces by using the framework of generalised Narain lattices of Drinfel'd doubles \cite{Klimcik:1996nq}, see also \cite{Demulder:2023vlo}.

\paragraph{Non-geometric backgrounds and dualities.} In this paper we have examined the Ricci Flow Conjecture (RFC) in the context of \textit{geometric} flux-supported internal spaces. A natural extension arises when considering T-duality transformations in flux compactifications, which can lead to \textit{non-geometric} backgrounds. This prompts the intriguing question of how backgrounds with non-geometric fluxes fit within the Swampland or the Landscape.
While this has been investigated in connection with various Swampland conjectures \cite{CaboBizet:2019sku,Gkountoumis:2023fym,Demulder:2023vlo,Rajaguru:2024emw,Becker:2024ijy}, it remains largely exploratory. Similarly, one would like to understand if an extension of the RFC may help identify the viability of non-geometric fluxes in low energy effective description of quantum gravity.
Non-geometric spaces are however by nature not smooth manifolds, rendering the question of obtaining a smooth flow rather ill-defined. However as already mentioned, reformulating generalised Ricci flow using generalised geometry, admits a T-dual covariant formulation \cite{Garcia-Fernandez:2016ofz}, see also \cite{Severa:2018pag,Streets:2024rfo}. Indeed, the flow equations are repackaged into the flow of the generalised metric and sourced by the generalised Ricci curvature. This is a direct manifestation of the action of (generalised) T-duality on the flow: while preserving the form of the flow equations, it maps solutions to solutions. One could thus contemplate flowing non-geometric backgrounds, albeit indirectly. By going to a non-geometric frame, the background is effectively described by a \textit{smooth} metric and a $\beta$-field. The resulting flow, once unpacked, would be a priori new and admit as a gradient function $\beta$-gravity functional \cite{Andriot:2012an,Andriot:2013xca} together with an corresponding   volume constraint.

The generalised Ricci is in addition not only preserved by T-duality but also its generalisation known as Poisson-Lie T-duality \cite{Severa:2016lwc}, see also the recent paper \cite{Streets:2024rfo}. This offers the possibility to explore a much wider set of supergravity backgrounds supported by geometric and non-geometric fluxes.

\paragraph{An information theoretic distance.} As extensively discussed in the main text, the proposed length is not a genuine distance. More notably, by missing the properties of symmetry and by not verifying the triangle inequality. Properties which, in other approaches \cite{Mohseni:2024njl,Calderon-Infante:2020dhm,Debusschere:2024rmi}, are also found to be absent. These observations may be a strong indication that the formulation of the SDC in presence of potentials, or equivalently that of the RFC with or without fluxes, should be formulated in terms of some sorts of cost rather than distance. Ricci flow and the generalised Ricci flow however admit an alternative description which comes with a natural distance measure. The authors in \cite{mccann2010ricci} showed that Ricci flow can be directly related to the theory of optimal transport. These results were later extended to the generalised Ricci flow in \cite{Kopfer2024}. The latter describes the most efficient way to transform one distribution into another, given a cost function. However, the optimal transport cost leads to a genuine distance: the Wasserstein distance. It would be interesting to explore this further and could help reinstate a proper distance within the Swampland Ricci Flow Conjecture. In addition, this would provide a novel window to explore the Swampland using classical information theoretical tools, and possibly make contact with the conclusions of \cite{Stout:2021ubb,Stout:2022phm}.

\subsubsection*{Acknowledgements}
We thank  Chris Blair, Davide De Biasio, Alessandra Gnecchi, Thomas Grimm, Christian Knei\ss l, Yixuan Li, Joaquin Masias, Carmine Montella, Georges Obied, Fridrich Valach, Irene Valenzuela for useful discussions. The work of SD and DL is supported by the Origins Excellence Cluster and by the German-Israel-Project (DIP) on Holography and the Swampland.
SD is also supported by an Azrieli fellowship funded by the Azrieli foundation, by the Israel Science Foundation (grant No. 1417/21),  by Carole and Marcus Weinstein through the BGU Presidential Faculty Recruitment Fund, by the ISF Center of Excellence for theoretical high energy physics, and by the ERC Starting Grant dSHologQI (project number 101117338). 
SD would like to thank the Max-Planck Gesellschaft, through ``The Max Planck-Israel Program'', and the Max-Planck Institute for Physics for support and hospitality during the end stages of this project. DL and TR would like to thank the Erwin Schr\"odinger International Institute for Mathematics and Physics for hospitality and the stimulating discussions during the workshop ``The Landscape vs. the Swampland''.

\newpage
\appendix
\addtocontents{toc}{\protect\setcounter{tocdepth}{1}}
\section{Relation to renormalisation group flow}\label{app:RG-flow}

As mentioned previously, there is a close connection between RG flow and (generalised) Ricci flow. The NSNS sector of the string effective action can be understood as the low-energy effective theory of closed strings on a curved background, described by a non-linear $\sigma$-model with target space parametrize by the triple $\{g,B,\phi\}$. The resulting RG $\beta$-functions or Weyl anomaly coefficients\footnote{There is a subtle difference between these two quantities that is essentially related to the difference between scale and conformal invariance. We refer to \cite{Witten:2024yod} and references therein.} then coincide with the equations of motion of the string effective action. Let us illustrate the connection on an explicit example. Take the bosonic $\sigma$-model with target space $S^3$ (equivalently the $SU(2)$ principal chiral model)
\begin{align}
    S &= \frac{1}{4\pi}\int \mathrm{d}^2\sigma \,g_{ij} \partial_\mu X^i \partial^\mu X^j
\end{align}
and write  $g_{ij}=r^2 g^0_{ij}$ with $r$ radius of the $S^3$. We can think of  $r^{-1}$ as coupling constant of our theory, i.e. we identify $g^2 \sim 1/r^2$. Now it is well known that the RG $\beta$-functions of the bosonic $\sigma$-model are given by $\beta_{ij}(g) = R_{ij}(g)$ such that 
\begin{align}
    \frac{\partial}{\partial\log(\Lambda)} g_{ij} = \beta_{ij}(g) = R_{ij}(g)\,,
\end{align}
where $\Lambda$ is the RG scale. We can recover standard Ricci flow by introducing RG time via $\Lambda = \Lambda_0 e^{-2t}$ such that the equation becomes
\begin{align}
     \frac{\partial}{\partial t} g_{ij} =  -2R_{ij}(g)\,.
\end{align}
Therefore we see that $\Lambda$ and $t$ are inversely related: while $\Lambda \to +\infty$ corresponds to going towards the UV, $t \to \infty$ describes a flow towards the IR.
In particular looking at the (Ricci) flow equation of the radius for the sphere  \eqref{eq:flow_sphere}, we see that in the IR ($t \to +\infty$) the radius vanishes, while in the UV ($t \to - \infty$) the sphere decompactions. Since the coupling scales with the inverse radius, $g \sim 1/r$, we recover the well known fact, that the Principal Chiral Model is asymptotically free.

\section{Generalised geometry of generalised Ricci flow}\label{app:GG_and_T-duality_form}
A crucial property of generalised Ricci flow is that T-duality maps solutions of the flow to other solutions, see \cite{garcia2021generalized} for an extensive discussion. Note that this is in contrast to conventional Ricci flow, which is not closed under T-duality since, T-duality generically mixes metric and Kalb-Ramond components. Generalised Ricci flow thus admits a formulation or repackaging in using generalised geometry. Define the generalised metric
\begin{align}\label{eq:gen_m}
    \mathcal H=\begin{pmatrix}
        g-Bg^{-1}B & \;Bg^{-1}\\
        -g^{-1}B & \;g
    \end{pmatrix}\,,
\end{align}
Then the flow in terms of the generalised metric for given background fields $(G,B)$ becomes \begin{align}
    \mathcal G^{-1}\frac{\partial}{\partial t}\mathcal G=-2\mathcal{R}c(\mathcal G)\,,
\end{align}
where  $\mathcal G=\eta\mathcal H$, with $\eta$ the $O(D, D)$ invariant metric and the indices have been suppressed, and the generalised Ricci tensor is given by
\begin{align}
    \mathcal R c= \begin{pmatrix}
		g^{-1}R+\tfrac{1}{2}g^{-1}H^2 & \tfrac{1}{2}g^{-1}\mathrm d^\ast Hg^{-1}\\
		-\tfrac{1}{2}\mathrm d^\ast H & -R\,g^{-1}-\tfrac{1}{2}H^2g^{-1}
	\end{pmatrix}\, .
\end{align}
A convenient alternative form of the flow is, and reinstating the indices
\begin{align}\label{eq:flow_gb}
	\frac{\partial}{\partial t}(G-B)_{ij}(X,Y)=-2R_{ij}^+(X,Y)\,,
\end{align}
here $R_{ij}^+$ is the Bismut-Ricci tensor associated to the Bismut connection $\nabla^+=\nabla+\tfrac{1}{2}g^{-1}H$ and takes the form
\begin{align}\label{eq:BismutRicci}
	R_{ij}^+=R_{ij}-\tfrac{1}{4}(H^2)_{ij}-\tfrac{1}{2}(\mathrm d^\ast H)_{ij}\,,
\end{align}
where $(H^2)_{ij}=H_{ikl}H_{j}{}^{kla}$. Including the dilaton field $\phi$, this becomes twisted Bakry-Emery curvature
\begin{align}
    R_{ij}^{H,\phi}= R_{ij}-\frac{1}{4}(H^2)_{ij}+2\nabla_i\nabla_j \phi-\frac{1}{2}(\mathrm d^\star H +\imath_{2\nabla \phi}H)_{ij}\,.
\end{align}
The associated generalised scalar curvature is
\begin{align}\label{eq:scalarcurv}
   S(g,H,\phi)=\left(R-\frac{1}{12}|H|^2+4\Delta f-4|\nabla \phi|^2 \right)\,.
\end{align}

\section{Gradient flow of $\mathcal{S}_{\mathrm{EH}}$ and modified flow equations for $f$}\label{app:modified_flows}

\subsection{Variation of $\mathcal{S}_{\mathrm{EH}}$}
We start from the Einstein Hilbert functional in the presence of fluxes, which we denote $\mathcal{S}_{\mathrm{EH}}$. Taking the one parameter variation of $\mathcal{S}_{\mathrm{EH}}$ with respect to $t$ gives (\cite{Oliynyk:2005ak} or \cite{garcia2021generalized} Lemma 6.7 for details)
\begin{align}\label{eq:variation_S_H}
    \frac{\mathrm d}{ \mathrm d t} \mathcal{S}_{\mathrm{EH}} &=\frac{\mathrm d}{\mathrm d t} \int_K d V_g\,e^{-f}\left(R-\frac{1}{12}|H|^2+|\nabla f| \right)\nonumber\\
    &= \int_K d V_g\,e^{-f} \Bigg[\left(-R^{ij}-\nabla^i \nabla^j f + \frac{1}{4} H^i_{kl}H^{jkl}\right) \frac{\partial g_{ij}}{\partial t}\nonumber\\
    &\qquad + \left(R-\frac{1}{12}|H|^2+2 \Delta f -|\nabla f|^2 \right)\left(\frac{1}{2}g^{ij}\frac{\partial g_{ij}}{\partial t}-\frac{\partial f}{\partial t} \right)\nonumber\\
    &\qquad + \frac{1}{2}\left(\nabla_k H^{kij}-H^{kij}\nabla_k f \right) \frac{\partial B_{ij}}{\partial t}\Biggr]\,.
\end{align}
Now recall that the $\mathcal F$ functional is identical to $\mathcal{S}_{\mathrm{EH}}$ but with the additional constrain of $f$ being subject to the constraint $\int \mu_t\equiv \int e^{-f}\mathrm{d}V=1$. In fact it is easy to see that this constraint implies that
\begin{align}\label{eq:unit_volume_implication}
\left(\frac{1}{2}g^{ij}\frac{\partial g_{ij}}{\partial t}-\frac{\partial f}{\partial t} \right) = 0
\end{align}
and therefore from the above calculation it follows that $\mathcal{F}$\footnote{The gradient is with respect to the basis $(\frac{\partial}{\partial g_{ij}},\frac{\partial}{\partial B_{ij}})$.} defines a gradient flow, where the flow equations can be read off from \eqref{eq:variation_S_H} and are nothing else than the equations of motion for $g,B$. The resulting flow equation for $f$ can be computed in turn from \eqref{eq:unit_volume_implication} and reads
\begin{align}
    \frac{\partial}{\partial t} f = - \Delta f - R + \frac{1}{4} |H|^2\,.
\end{align}
Furthermore we see that along the flow, i.e when the flow equations are fulfilled the variation of $\mathcal{F}$ can be written as
\begin{align}
     \frac{\mathrm d}{ \mathrm d t} \mathcal{F}(g,H,f)
    = \int_K d V_g\,e^{-f} \Bigg[ \left|-R^{ij}-\nabla^i \nabla^j f + \frac{1}{4} H^i_{kl}H^{jkl}\right|^2\nonumber + \left| \frac{1}{2}\left(\nabla_k H^{kij}-H^{kij}\nabla_k f \right)\right|^2 \Biggr] \geq 0
\end{align}
and therefore is monotonous along the flow.\footnote{Actually $\mathcal{F}$ is strictly monotonous along the flow, since by definition the summands are only zero at the fixed points of the flow.}
\subsection{Modified flow equations for $f$}\label{sec:modified_flows}
As discussed in section \ref{sec:role_f}, the functional $\mathcal{S}_{\mathrm{EH}}$ stays monotonous even without imposing the unit volume constraint, if we instead impose a modified flow equation for $f$. In fact from the variation above we see that the variation is manifestly positive if each line can be written as square of terms. In fact we would like to write
\begin{align}\label{eq:term_from_variation}
\left(R-\frac{1}{12}|H|^2+2 \Delta f -|\nabla f|^2 \right)\left(\frac{1}{2}g^{ij}\frac{\partial g_{ij}}{\partial t}-\frac{\partial f}{\partial t} \right) = S(g,B,f) \cdot \frac{\partial_t\mu_t}{\mu_t}
\end{align}
as a square. Taking the evolution of $\mu_t$ to be given by
\begin{align}\label{eq:flow_mu}
    \frac{\partial}{\partial t} \log{\mu} = \alpha \, S(g,B,f)
\end{align}
with $\alpha$ a yet undetermined integer
the variation of the functional $\mathcal{S}_{\mathrm{EH}}$ can then be written as a sum of quadratic terms
\begin{align}\label{eq:monotnous_function_alpha}
    \frac{\mathrm d}{\mathrm{d} t} \mathcal{S}_{\mathrm{EH}}  \sim \int_M \left(\frac{1}{2}|\partial_t g|^2 +  \frac{1}{2}|\partial_t B|^2 + \frac{1}{\alpha}\left|\frac{\partial_t\mu}{\mu}\right|^2\right)e^{-f}\mathrm{d}V \geq 0\,,
\end{align}
and hence $\mathcal{S}_{\mathrm{EH}}$ is guaranteed to be \textit{striclty} monotonically increasing along the flow whenever we choose $\alpha$ greater or equal to $0$. Choosing $\alpha=0$ is of course nothing else than the standard constraint $\int \mu_t= 1$.

From eq. \eqref{eq:flow_mu} we can infer the flow of the scalar $f$ by using $\frac{\partial_t \mu}{\mu}= 1/2 g^{ij} \partial_t g_{ij}-\partial_t f$ and the definition of $S(g,B,f)$.
First note that tracing the flow equation $\partial_t g_{ij}$ we get
\begin{align}
    \frac{1}{2}g^{ij} \partial_t g_{ij} = - (R+\Delta f - \frac{1}{4} |H|^2)
\end{align}
such that
\begin{align}\label{eq:modified_flow_f}
    0 &= \alpha\, S - \partial_t \log(\mu)\nonumber\\
    &=\alpha \left(R-\frac{1}{12}|H|^2+2 \Delta f -|\nabla f|^2  \right)+ (R+\Delta f - \frac{1}{4} |H|^2)+\partial_t f\nonumber\\
    &=(\alpha + 1) R - \frac{\alpha+3}{12} |H|^2 + (2\alpha+1)\Delta f - \alpha |\nabla f|^2+\partial_t f\,.
\end{align}
\begin{align}
    \frac{\partial}{\partial t} f = T^{(\alpha)}_f \equiv -(\alpha + 1) R + \frac{\alpha+3}{12} |H|^2 - (2\alpha+1)\Delta f + \alpha |\nabla f|^2 
\end{align}
As a consistency check we can set $\alpha=0$, immediately implying the $\int\mu_t=1$ constraint and we indeed get back the correct formula for $\partial_t f$.  For example choosing $\alpha=2$ in analogy to the other flow equations we obtain
\begin{align}
    -\frac{\partial}{\partial t}  f = 3R -\frac{5}{12}|H|^2 + 5\Delta f -2 |\nabla f|^2\,.
\end{align}
This is not the beta-function of the dilaton. Note nevertheless that  whenever we hit a fixed point of the flow, we have
\begin{gather}
\begin{aligned}
    0&= \frac{\partial}{\partial t}  g_{ij} =-R_{ij}-\nabla_i \nabla_j f + \frac{1}{4} H_{ikl}H_i^{\ kl}\,,\\
    0&= -\frac{\partial}{\partial t}  f = 3R -\frac{5}{12}|H|^2 + 3\Delta f -2 |\nabla f|^2\,.
\end{aligned}
\end{gather}
Taking the trace of the first equation an substituting into the dilaton one then we obtain
\begin{align}
    0= \frac{1}{3}|H|^2 + 2\Delta f - 2|\nabla f|^2 \,\equiv \beta^f\,,
\end{align}
which is nothing else (upon identifying $f=2\phi$) than the standard dilaton equation of motion 
\begin{align}
0=|\nabla \phi|^2-\frac{1}{2}\Delta \phi - \frac{1}{24}|H|^2\,,
\end{align}
and hence at the fixed point the dilaton equation of motion is also satisfied. Finally observe that in principle we can also choose $\alpha=-1$ to obtain
\begin{align}
    \frac{\partial}{\partial t} f = \frac{1}{6}|H|^2 + \Delta f - |\nabla f|^2 \,\sim \beta^f\,.
\end{align}
and hence also $f$ evolves according to the NLSM $\beta$-function and it is justified to identify $f\equiv 2\phi$. However in this case the functional $\mathcal{S}_{\mathrm{EH}}$ is generically no longer monotonic along the flow, since the last quadratic term will appear with a minus in front. 

\paragraph{Examples of modified flow}
Let us once more look at the canonical example of $S^3$ with $H$-flux, for different values of $\alpha$.

\begin{figure}[t]
	\centering
	\includegraphics[scale=0.5]{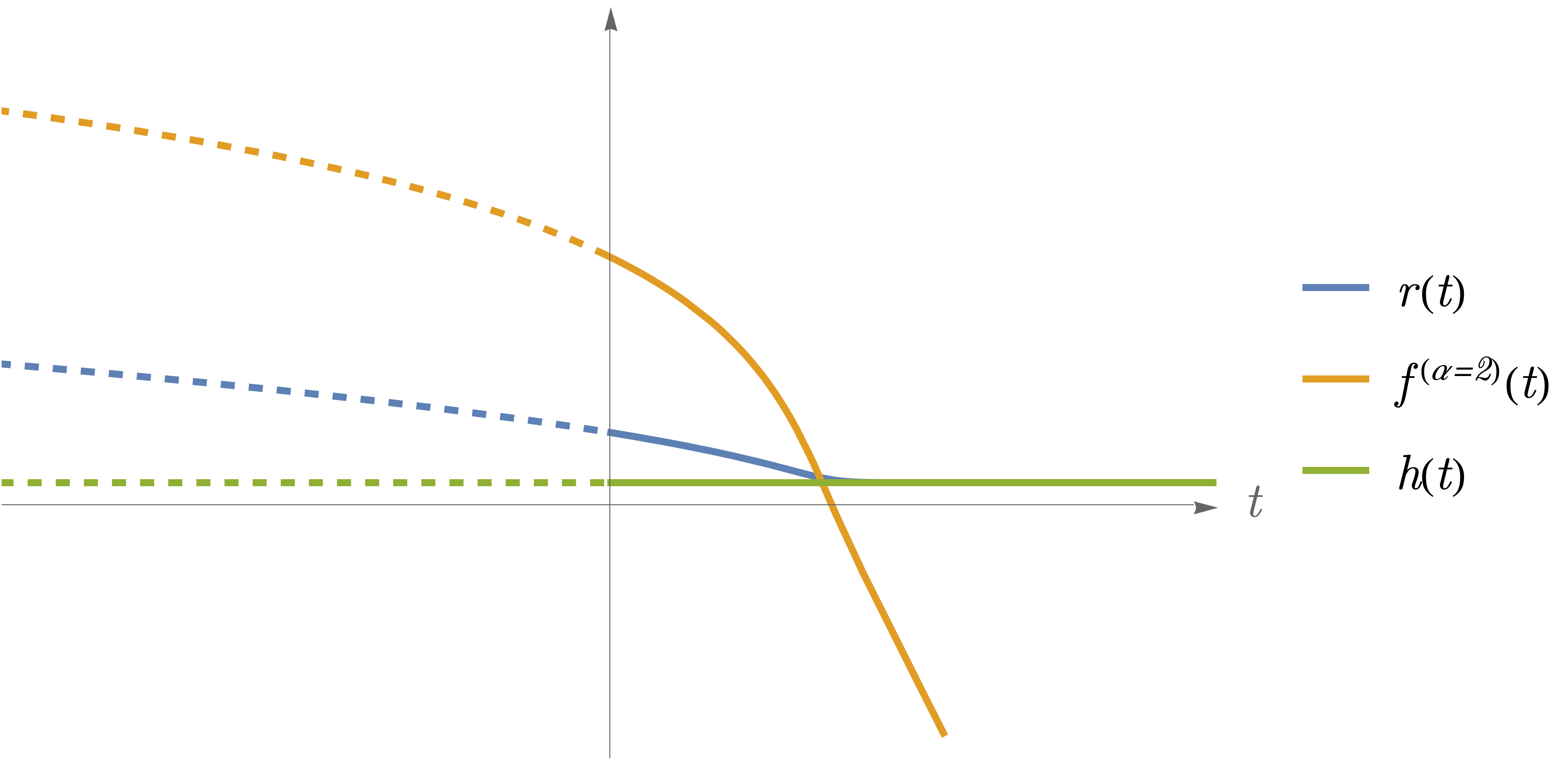}
    \includegraphics[scale=0.5]{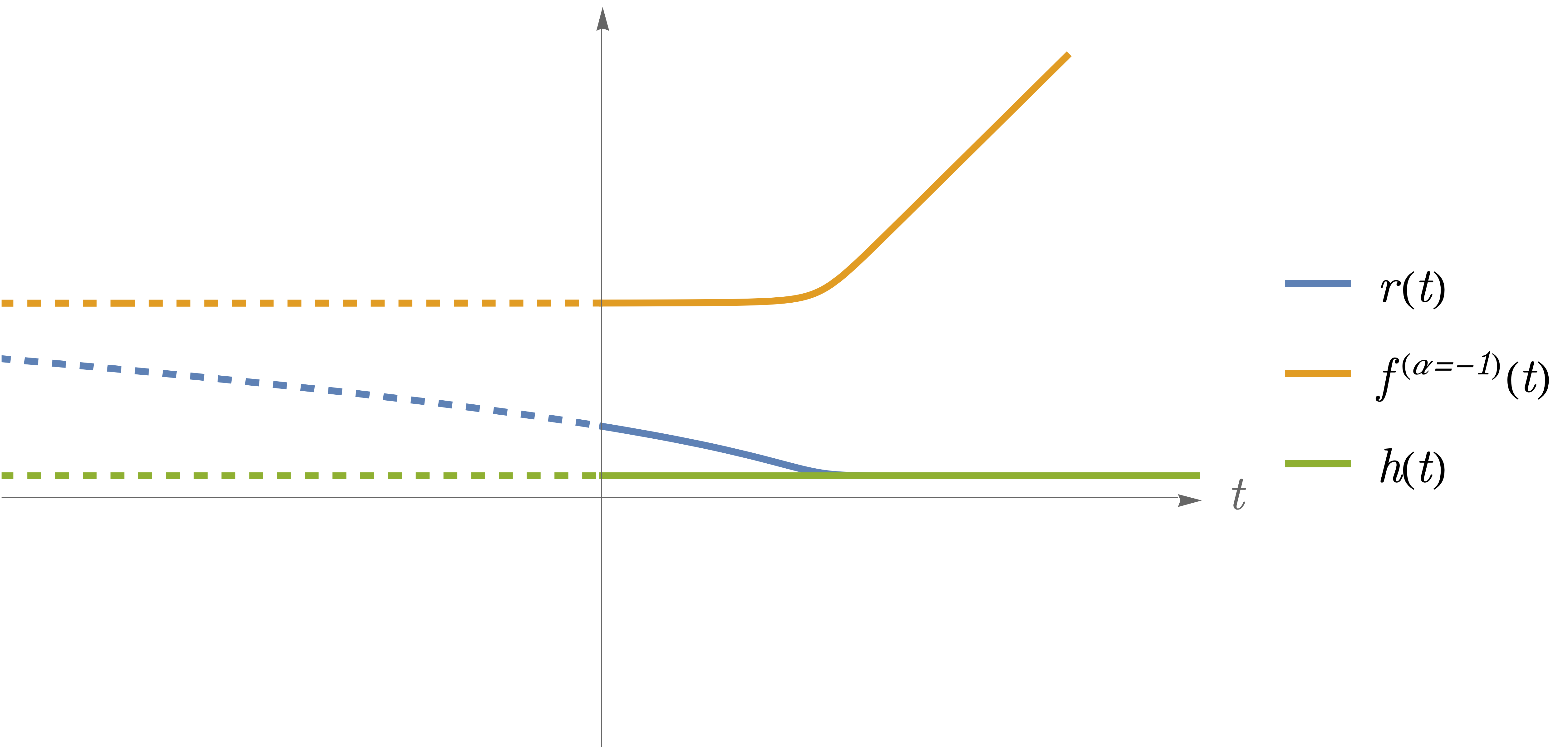}
	\caption{Plotted is the behaviour of the modified flows described in this section on the example of $S^3$ with $h=1$ units of flux. It is clear that although $r$ stabilises, $f$ diverges towards $- \infty$ for the modified flow with $\alpha=2$ (left plot) while it diverges to $+\infty$ for $\alpha=-1$ (right plot).}\label{fig:S3_H_with_fmod_backwards}
\end{figure}

For $\alpha=2$ we obtain 
\begin{align}
    \frac{\partial}{\partial t} f^{(\alpha=2)} = T^{(\alpha=2)}_f = -\frac{18}{r(t)^2}+\frac{10 h^2}{r(t)^6}
\end{align}
Since $r(t)\to \sqrt{h}$  for $t \to \infty$ (as determined by the flow equation for the metric $g$), $\partial_f$ is not vanishing but approaching the constant value $\frac{-8}{h}$ and therefore $f(t)$ is diverging towards $- \infty$. Therefore the combined flow $\{g(t),f(t)\}$ does not reach a fixed point as $t \to \infty (r(t) \to \sqrt{k})$. However similar to the standard case, starting at $r(0)>\sqrt{h}$, going to negative $t$ there is a fixed point at $t \to - \infty$ where $r(t) \to \infty$, cf. figure \ref{fig:S3_H_with_fmod_backwards} left panel.

Choosing on the other hand $\alpha=-1$ we obtain
\begin{align}
    \frac{\partial}{\partial t} f^{(\alpha=-1)} = T^{(\alpha=-1)}_f = \frac{4 h^2}{r(t)^6}
\end{align}
This flow is illustrated in the right panel of figure \ref{fig:S3_H_with_fmod_backwards}.

\subsubsection{Including the dilaton explicitly}\label{sec:incl_dilaton}

In the last paragraph we briefly mentioned that upon setting $\alpha=-1$  in the modified flow equation for $f$, the resulting flow equation is exactly the dilaton equation of motionon. Hence one can identify $f\equiv 2\phi$ with a dilaton. This is indeed an intriguing feature since it seems interesting to have a flow for the dilaton according to its $\beta$-function given we already have $g,B$ evolving according to their $\beta$-functions. Concretely we would like to have a system of flow equations for $g,B$ and a scalar field $\phi$ that we identify with the dilaton that reads\footnote{This is indeed a well defined system of PDE, since by applying a diffeomorphism we can decouple $\phi$ from the equations of $g,B$. The remaining equation is a Heat-like equation and can be solved a posteriori.}
\begin{gather}
    \begin{aligned}
    \frac{\partial}{\partial t}g_{ij}&=-2\beta^g_{ij} =-2 R_{ij}+\frac{1}{2}H^2_{ij} -4\nabla_i \nabla_j \phi\,,\\
    \frac{\partial}{\partial t}B_{ij}&=-2\beta^B_{ij} =\nabla^k H_{kij}-2H_{kij}\nabla^k\phi\,,\\
    \frac{\partial}{\partial t}\phi&=-2\beta^\phi =\Delta \phi -  2(\nabla \phi)^2 +\frac{1}{12}H^2\,.
\end{aligned}
\end{gather}
\noindent Trying to directly identify $f=2 \phi$  leads back to the discussion of the last paragraph and is therefore not desirable. However it has been realised in \cite{streets2022scalarcurvatureentropygeneralized}
that the system above can be realised as (the gauge fixed) gradient flow of the functional 
\begin{gather}
    \begin{aligned}\label{eq:functionl_dilaton_NSNS}
    \mathcal F(g,H,f=2\phi+\tilde{f})&=\int\left(R-\frac{1}{12}|H|^2+|\nabla (2\phi+\tilde{f})|^2 \right)e^{-2\phi}e^{-\tilde{f}}\mathrm d V_g\\
    &=\int\left(\left(R-\frac{1}{12}|H|^2+4|\nabla \phi|^2 \right)+4\nabla \phi \nabla \tilde{f} + (\nabla \tilde{f})^2\right)e^{-2\phi}e^{-\tilde{f}}\mathrm d V_g\,,
\end{aligned}
\end{gather}
supplemented by the constraint 
\begin{align}
    \int \mu^\phi_t = \int e^{-2\phi}e^{-\tilde{f}}\mathrm{d}V_g = 1.
\end{align}
The flow equation for $\tilde{f}$ can be again calculated from this constraint and reads\footnote{After performing the diffeomorphism that decouples $f$ form the other flow equations via $\mathcal{L}_{\nabla \tilde{f}}$. }
\begin{align}
   \frac{ \partial}{\partial t} \tilde{f}= \tilde{T}_{\tilde{f}} \equiv  - \Delta \tilde{f} - 2 \nabla \phi \nabla \tilde{f} - R + \frac{1}{12}H^2 - 2 \Delta \phi + (\nabla \tilde{f})^2.
\end{align}
It has to be stressed that although the resulting flow is gradient, it is only the flow equations for $g$ and $B$ that appear as the gradient of $\mathcal{F}(g,H,2\phi+\tilde{f})$. The dilaton equation is introduced in a somewhat ``artificial" manner and therefore also enters the distance only indirectly via the other flow equations.

\paragraph{Example including $\phi$}
We look again at $S^3$ with $H$-flux. Then we obtain (additionally to the equation for $r(t)$)
\begin{gather}
\begin{aligned}
    \frac{ \partial}{\partial t} \phi(t) &=\beta^\phi= -\frac{2h^2}{r(t)^6}\,,\\
    \frac{ \partial}{\partial t} \tilde{f}(t) &= \tilde{T}_{\tilde{f}} = -\frac{6}{r(t)^2}+\frac{2 h^2}{r(t)^6}\,.
\end{aligned}
\end{gather}
\noindent Again we encounter a situation in which even when $r(t)$ approaches its fixed point value $r=\sqrt{h}$, $\phi$ and $\tilde{f}$ do not approach a constant. However they diverge in a way such that the generalised volume constraint $\int_K \mu^\phi_t=1$ is fulfilled.

\begin{figure}[t]
	\centering
	\includegraphics[scale=0.55]{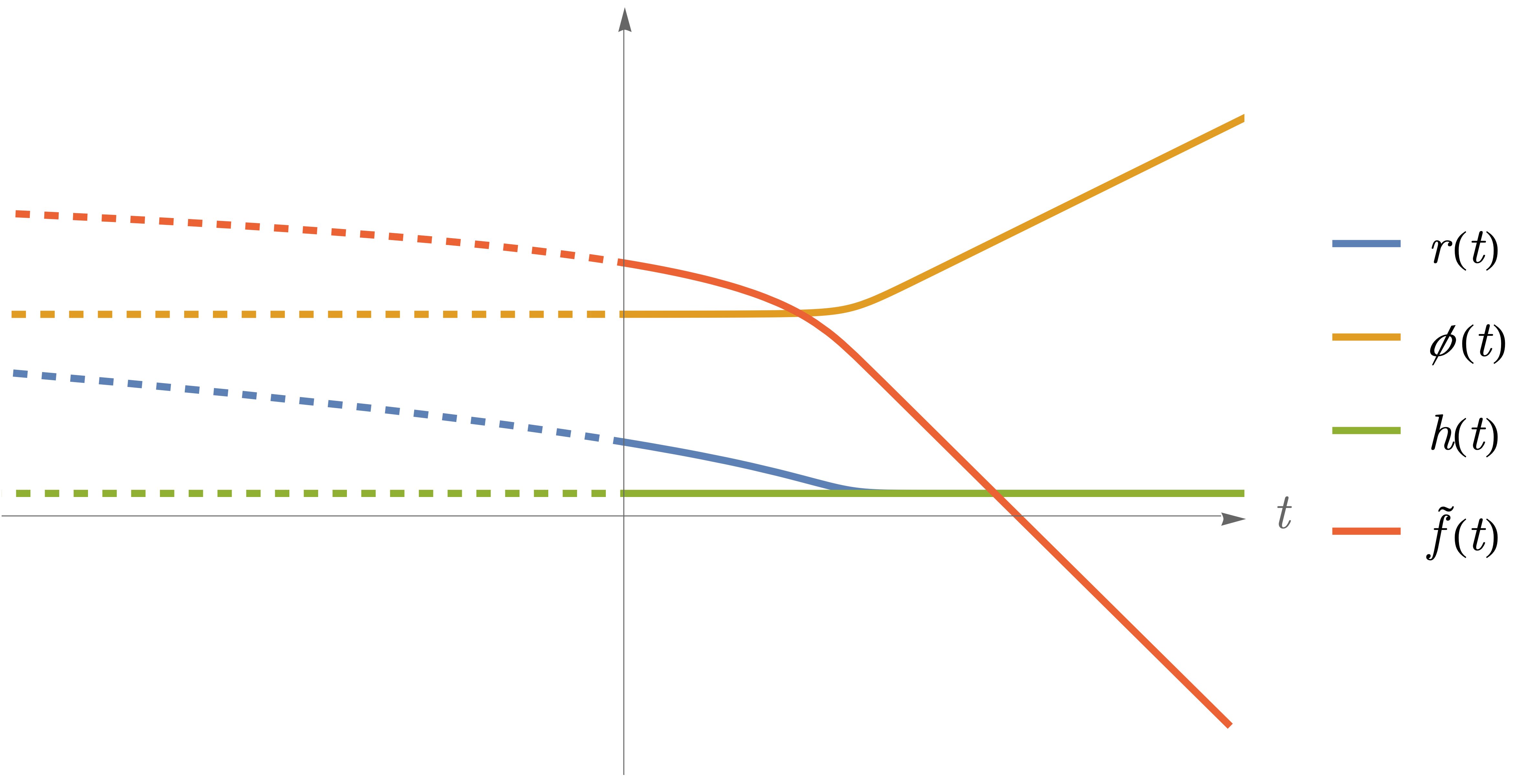}
	\caption{Illustrated is the flow behaviour of the modified dlow including the dilaton via the split $f=2\phi+\tilde{f}$. Both $\phi$ and $\tilde{f}$ diverge in a way such that the unit volume constraint is anyway satisfied.}\label{fig:S3_H_with_phi_backwards}
\end{figure}

\bibliographystyle{JHEP}
\bibliography{bib_gen_ricci_conj}

\providecommand{\href}[2]{#2}\begingroup\raggedright\begin{thebibliography}{10}

\bibitem{Vafa:2005ui}
C.~Vafa, \emph{{The String landscape and the swampland}},  \href{http://arxiv.org/abs/hep-th/0509212}{{\tt hep-th/0509212}}.

\bibitem{Ooguri:2004zv}
H.~Ooguri, A.~Strominger and C.~Vafa, \emph{{Black hole attractors and the topological string}}, \href{http://dx.doi.org/10.1103/PhysRevD.70.106007}{\emph{Phys. Rev. D} {\bf 70} (2004) 106007}, [\href{http://arxiv.org/abs/hep-th/0405146}{{\tt hep-th/0405146}}].

\bibitem{Lust:2019zwm}
D.~L\"ust, E.~Palti and C.~Vafa, \emph{{AdS and the Swampland}}, \href{http://dx.doi.org/10.1016/j.physletb.2019.134867}{\emph{Phys. Lett. B} {\bf 797} (2019) 134867}, [\href{http://arxiv.org/abs/1906.05225}{{\tt 1906.05225}}].

\bibitem{Headrick:2006ti}
M.~Headrick and T.~Wiseman, \emph{{Ricci flow and black holes}}, \href{http://dx.doi.org/10.1088/0264-9381/23/23/006}{\emph{Class. Quant. Grav.} {\bf 23} (2006) 6683--6708}, [\href{http://arxiv.org/abs/hep-th/0606086}{{\tt hep-th/0606086}}].

\bibitem{DeWitt:1967yk}
B.~S. DeWitt, \emph{{Quantum Theory of Gravity. 1. The Canonical Theory}}, \href{http://dx.doi.org/10.1103/PhysRev.160.1113}{\emph{Phys. Rev.} {\bf 160} (1967) 1113--1148}.

\bibitem{Calderon-Infante:2020dhm}
J.~Calder\'on-Infante, A.~M. Uranga and I.~Valenzuela, \emph{{The Convex Hull Swampland Distance Conjecture and Bounds on Non-geodesics}}, \href{http://dx.doi.org/10.1007/JHEP03(2021)299}{\emph{JHEP} {\bf 03} (2021) 299}, [\href{http://arxiv.org/abs/2012.00034}{{\tt 2012.00034}}].

\bibitem{Demulder:2023vlo}
S.~Demulder, D.~L\"ust and T.~Raml, \emph{{Topology change and non-geometry at infinite distance}}, \href{http://dx.doi.org/10.1007/JHEP06(2024)079}{\emph{JHEP} {\bf 06} (2024) 079}, [\href{http://arxiv.org/abs/2312.07674}{{\tt 2312.07674}}].

\bibitem{Basile:2023rvm}
I.~Basile and C.~Montella, \emph{{Domain walls and distances in discrete landscapes}}, \href{http://dx.doi.org/10.1007/JHEP02(2024)227}{\emph{JHEP} {\bf 02} (2024) 227}, [\href{http://arxiv.org/abs/2309.04519}{{\tt 2309.04519}}].

\bibitem{Mohseni:2024njl}
A.~Mohseni, M.~Montero, C.~Vafa and I.~Valenzuela, \emph{{On Measuring Distances in the Quantum Gravity Landscape}},  \href{http://arxiv.org/abs/2407.02705}{{\tt 2407.02705}}.

\bibitem{Debusschere:2024rmi}
C.~Debusschere, F.~Tonioni and T.~Van~Riet, \emph{{A distance conjecture beyond moduli?}},  \href{http://arxiv.org/abs/2407.03715}{{\tt 2407.03715}}.

\bibitem{Hamilton1982}
R.~S. Hamilton, \emph{Three-manifolds with positive ricci curvature}, {\emph{Journal of Differential Geometry} {\bf 17} (1982) 255--306}.

\bibitem{perelman2002entropy}
G.~Perelman, \emph{The entropy formula for the ricci flow and its geometric applications}, {\emph{arXiv preprint math/0211159} (2002) }.

\bibitem{Perelman2003}
G.~Perelman, \emph{Ricci flow with surgery on three-manifolds}, {\emph{arXiv preprint math/0303109} (2003) }.

\bibitem{Oliynyk:2005ak}
T.~Oliynyk, V.~Suneeta and E.~Woolgar, \emph{{A Gradient flow for worldsheet nonlinear sigma models}}, \href{http://dx.doi.org/10.1016/j.nuclphysb.2006.01.036}{\emph{Nucl. Phys. B} {\bf 739} (2006) 441--458}, [\href{http://arxiv.org/abs/hep-th/0510239}{{\tt hep-th/0510239}}].

\bibitem{Tseytlin:2006ak}
A.~A. Tseytlin, \emph{{On sigma model RG flow, 'central charge' action and Perelman's entropy}}, \href{http://dx.doi.org/10.1103/PhysRevD.75.064024}{\emph{Phys. Rev. D} {\bf 75} (2007) 064024}, [\href{http://arxiv.org/abs/hep-th/0612296}{{\tt hep-th/0612296}}].

\bibitem{Papadopoulos:2024uvi}
G.~Papadopoulos and E.~Witten, \emph{{Scale and Conformal Invariance in 2d Sigma Models, with an Application to N=4 Supersymmetry}},  \href{http://arxiv.org/abs/2404.19526}{{\tt 2404.19526}}.

\bibitem{Papadopoulos:2024tgs}
G.~Papadopoulos, \emph{{Scale and Conformal Invariance in Heterotic $\sigma$-Models}},  \href{http://arxiv.org/abs/2409.01818}{{\tt 2409.01818}}.

\bibitem{DeBiasio:2022nsd}
D.~De~Biasio, J.~Freigang, D.~L\"ust and T.~Wiseman, \emph{{Gradient flow of Einstein-Maxwell theory and Reissner-Nordstr\"om black holes}}, \href{http://dx.doi.org/10.1007/JHEP03(2023)074}{\emph{JHEP} {\bf 03} (2023) 074}, [\href{http://arxiv.org/abs/2210.14705}{{\tt 2210.14705}}].

\bibitem{Kehagias:2019akr}
A.~Kehagias, D.~L\"ust and S.~L\"ust, \emph{{Swampland, Gradient Flow and Infinite Distance}}, \href{http://dx.doi.org/10.1007/JHEP04(2020)170}{\emph{JHEP} {\bf 04} (2020) 170}, [\href{http://arxiv.org/abs/1910.00453}{{\tt 1910.00453}}].

\bibitem{DeBiasio:2020xkv}
D.~De~Biasio and D.~L\"ust, \emph{{Geometric Flow Equations for Schwarzschild-AdS Space-Time and Hawking-Page Phase Transition}}, \href{http://dx.doi.org/10.1002/prop.202000053}{\emph{Fortsch. Phys.} {\bf 68} (2020) 2000053}, [\href{http://arxiv.org/abs/2006.03076}{{\tt 2006.03076}}].

\bibitem{Velazquez:2022eco}
D.~M. Vel\'azquez, D.~De~Biasio and D.~L\"ust, \emph{{Cobordism, singularities and the Ricci flow conjecture}}, \href{http://dx.doi.org/10.1007/JHEP01(2023)126}{\emph{JHEP} {\bf 01} (2023) 126}, [\href{http://arxiv.org/abs/2209.10297}{{\tt 2209.10297}}].

\bibitem{DeBiasio:2022zuh}
D.~De~Biasio, \emph{{On-Shell Flow}},  \href{http://arxiv.org/abs/2211.04231}{{\tt 2211.04231}}.

\bibitem{DeBiasio:2022omq}
D.~De~Biasio and D.~L\"ust, \emph{{Geometric flow of bubbles}}, \href{http://dx.doi.org/10.1016/j.nuclphysb.2022.115812}{\emph{Nucl. Phys. B} {\bf 980} (2022) 115812}, [\href{http://arxiv.org/abs/2201.01679}{{\tt 2201.01679}}].

\bibitem{garcia2021generalized}
M.~Garcia-Fernandez and J.~Streets, \emph{Generalized Ricci Flow}, vol.~76.
\newblock American Mathematical Soc., 2021.

\bibitem{Lee:2019wij}
S.-J. Lee, W.~Lerche and T.~Weigand, \emph{{Emergent strings from infinite distance limits}}, \href{http://dx.doi.org/10.1007/JHEP02(2022)190}{\emph{JHEP} {\bf 02} (2022) 190}, [\href{http://arxiv.org/abs/1910.01135}{{\tt 1910.01135}}].

\bibitem{streets2008regularity}
J.~Streets, \emph{Regularity and expanding entropy for connection ricci flow}, {\emph{Journal of Geometry and Physics} {\bf 58} (2008) 900--912}.

\bibitem{Garcia-Fernandez:2016ofz}
M.~Garcia-Fernandez, \emph{{Ricci flow, Killing spinors, and T-duality in generalized geometry}}, \href{http://dx.doi.org/10.1016/j.aim.2019.04.038}{\emph{Adv. Math.} {\bf 350} (2019) 1059--1108}, [\href{http://arxiv.org/abs/1611.08926}{{\tt 1611.08926}}].

\bibitem{streets2017generalized}
J.~Streets, \emph{Generalized geometry, t-duality, and renormalization group flow}, {\emph{Journal of Geometry and Physics} {\bf 114} (2017) 506--522}.

\bibitem{Candelas:1990pi}
P.~Candelas and X.~de~la Ossa, \emph{{Moduli Space of {Calabi-Yau} Manifolds}}, \href{http://dx.doi.org/10.1016/0550-3213(91)90122-E}{\emph{Nucl. Phys. B} {\bf 355} (1991) 455--481}.

\bibitem{Li:2023gtt}
Y.~Li, E.~Palti and N.~Petri, \emph{{Towards AdS distances in string theory}}, \href{http://dx.doi.org/10.1007/JHEP08(2023)210}{\emph{JHEP} {\bf 08} (2023) 210}, [\href{http://arxiv.org/abs/2306.02026}{{\tt 2306.02026}}].

\bibitem{Shiu:2023bay}
G.~Shiu, F.~Tonioni, V.~Van~Hemelryck and T.~Van~Riet, \emph{{Connecting flux vacua through scalar field excursions}}, \href{http://dx.doi.org/10.1103/PhysRevD.109.066017}{\emph{Phys. Rev. D} {\bf 109} (2024) 066017}, [\href{http://arxiv.org/abs/2311.10828}{{\tt 2311.10828}}].

\bibitem{Palti:2024voy}
E.~Palti and N.~Petri, \emph{{A positive metric over DGKT vacua}}, \href{http://dx.doi.org/10.1007/JHEP06(2024)019}{\emph{JHEP} {\bf 06} (2024) 019}, [\href{http://arxiv.org/abs/2405.01084}{{\tt 2405.01084}}].

\bibitem{streets2023scalar}
J.~Streets, \emph{Scalar curvature, entropy, and generalized ricci flow}, {\emph{International Mathematics Research Notices} {\bf 2023} (2023) 9481--9510}, [\href{http://arxiv.org/abs/2207.13197}{{\tt 2207.13197}}].

\bibitem{Kehagias:2022mik}
A.~Kehagias, H.~Partouche and N.~Toumbas, \emph{{A unimodular-like string effective description}}, \href{http://dx.doi.org/10.1016/j.nuclphysb.2023.116196}{\emph{Nucl. Phys. B} {\bf 991} (2023) 116196}, [\href{http://arxiv.org/abs/2212.14659}{{\tt 2212.14659}}].

\bibitem{kotschwar2009backwardsuniquenessricciflow}
B.~Kotschwar, \emph{Backwards uniqueness of the ricci flow},  \href{http://arxiv.org/abs/0906.4920}{{\tt 0906.4920}}.

\bibitem{Severa:2018pag}
P.~\v{S}evera and F.~Valach, \emph{{Courant Algebroids, Poisson\textendash{}Lie T-Duality, and Type II Supergravities}}, \href{http://dx.doi.org/10.1007/s00220-020-03736-x}{\emph{Commun. Math. Phys.} {\bf 375} (2020) 307--344}, [\href{http://arxiv.org/abs/1810.07763}{{\tt 1810.07763}}].

\bibitem{Kopfer2024}
E.~Kopfer and J.~Streets, \emph{Optimal transport and generalized ricci flow}, \href{http://dx.doi.org/10.3842/sigma.2024.003}{\emph{Symmetry, Integrability and Geometry: Methods and Applications} (Jan., 2024) }.

\bibitem{hamilton1982inverse}
R.~S. Hamilton, \emph{The inverse function theorem of nash and moser}, .

\bibitem{lee2012smooth}
J.~M. Lee and J.~M. Lee, \emph{Smooth manifolds}.
\newblock Springer, 2012.

\bibitem{Fei:2018mzf}
T.~Fei, B.~Guo and D.~H. Phong, \emph{{Parabolic dimensional reductions of 11-dimensional supergravity}}, \href{http://dx.doi.org/10.2140/apde.2021.14.1333}{\emph{Anal. Part. Diff. Eq.} {\bf 14} (2021) 1333--1361}, [\href{http://arxiv.org/abs/1806.00583}{{\tt 1806.00583}}].

\bibitem{Fei:2021pbw}
T.~Fei and D.~H. Phong, \emph{{Symplectic geometric flows}}, \href{http://dx.doi.org/10.4310/PAMQ.2023.v19.n4.a6}{\emph{Pure Appl. Math. Quart.} {\bf 19} (2023) 1853--1871}, [\href{http://arxiv.org/abs/2111.14048}{{\tt 2111.14048}}].

\bibitem{DeBiasio:2023hzo}
D.~De~Biasio, \emph{{Geometric flows and the Swampland}},  \href{http://arxiv.org/abs/2307.08320}{{\tt 2307.08320}}.

\bibitem{Li:2021utg}
Y.~Li, \emph{{An Alliance in the Tripartite Conflict over Moduli Space}},  \href{http://arxiv.org/abs/2112.03281}{{\tt 2112.03281}}.

\bibitem{Bonnefoy:2019nzv}
Q.~Bonnefoy, L.~Ciambelli, D.~L\"ust and S.~L\"ust, \emph{{Infinite Black Hole Entropies at Infinite Distances and Tower of States}}, \href{http://dx.doi.org/10.1016/j.nuclphysb.2020.115112}{\emph{Nucl. Phys. B} {\bf 958} (2020) 115112}, [\href{http://arxiv.org/abs/1912.07453}{{\tt 1912.07453}}].

\bibitem{Cribiori:2022cho}
N.~Cribiori, M.~Dierigl, A.~Gnecchi, D.~L\"ust and M.~Scalisi, \emph{{Large and small non-extremal black holes, thermodynamic dualities, and the Swampland}}, \href{http://dx.doi.org/10.1007/JHEP10(2022)093}{\emph{JHEP} {\bf 10} (2022) 093}, [\href{http://arxiv.org/abs/2202.04657}{{\tt 2202.04657}}].

\bibitem{Cribiori:2022nke}
N.~Cribiori, D.~L\"ust and G.~Staudt, \emph{{Black hole entropy and moduli-dependent species scale}}, \href{http://dx.doi.org/10.1016/j.physletb.2023.138113}{\emph{Phys. Lett. B} {\bf 844} (2023) 138113}, [\href{http://arxiv.org/abs/2212.10286}{{\tt 2212.10286}}].

\bibitem{Delgado:2022dkz}
M.~Delgado, M.~Montero and C.~Vafa, \emph{{Black holes as probes of moduli space geometry}}, \href{http://dx.doi.org/10.1007/JHEP04(2023)045}{\emph{JHEP} {\bf 04} (2023) 045}, [\href{http://arxiv.org/abs/2212.08676}{{\tt 2212.08676}}].

\bibitem{Montero:2024qtz}
M.~Montero and I.~Valenzuela, \emph{{Quantum corrections to DGKT and the Weak Gravity Conjecture}},  \href{http://arxiv.org/abs/2412.00189}{{\tt 2412.00189}}.

\bibitem{Hoare:2019ark}
B.~Hoare, N.~Levine and A.~A. Tseytlin, \emph{{Integrable 2d sigma models: quantum corrections to geometry from RG flow}}, \href{http://dx.doi.org/10.1016/j.nuclphysb.2019.114798}{\emph{Nucl. Phys. B} {\bf 949} (2019) 114798}, [\href{http://arxiv.org/abs/1907.04737}{{\tt 1907.04737}}].

\bibitem{Grimm:2018ohb}
T.~W. Grimm, E.~Palti and I.~Valenzuela, \emph{{Infinite Distances in Field Space and Massless Towers of States}}, \href{http://dx.doi.org/10.1007/JHEP08(2018)143}{\emph{JHEP} {\bf 08} (2018) 143}, [\href{http://arxiv.org/abs/1802.08264}{{\tt 1802.08264}}].

\bibitem{Corvilain:2018lgw}
P.~Corvilain, T.~W. Grimm and I.~Valenzuela, \emph{{The Swampland Distance Conjecture for K\"ahler moduli}}, \href{http://dx.doi.org/10.1007/JHEP08(2019)075}{\emph{JHEP} {\bf 08} (2019) 075}, [\href{http://arxiv.org/abs/1812.07548}{{\tt 1812.07548}}].

\bibitem{cao1986deformation}
H.-D. Cao, \emph{Deformation of kahler metrics to kahler-einstein metrics on compact kahler manifolds},  tech. rep., Princeton Univ., NJ (USA), 1986.

\bibitem{apostolov2022generalized}
V.~Apostolov, X.~Fu, J.~Streets and Y.~Ustinovskiy, \emph{The generalized k$\backslash$" ahler calabi-yau problem}, {\emph{arXiv preprint arXiv:2211.09104} (2022) }.

\bibitem{Grana:2005ny}
M.~Grana, J.~Louis and D.~Waldram, \emph{{Hitchin functionals in N=2 supergravity}}, \href{http://dx.doi.org/10.1088/1126-6708/2006/01/008}{\emph{JHEP} {\bf 01} (2006) 008}, [\href{http://arxiv.org/abs/hep-th/0505264}{{\tt hep-th/0505264}}].

\bibitem{Benmachiche:2006df}
I.~Benmachiche and T.~W. Grimm, \emph{{Generalized N=1 orientifold compactifications and the Hitchin functionals}}, \href{http://dx.doi.org/10.1016/j.nuclphysb.2006.05.003}{\emph{Nucl. Phys. B} {\bf 748} (2006) 200--252}, [\href{http://arxiv.org/abs/hep-th/0602241}{{\tt hep-th/0602241}}].

\bibitem{Pestun:2005ni}
V.~Pestun, \emph{{Black hole entropy and topological strings on generalized CY manifolds}}, \href{http://dx.doi.org/10.1088/1126-6708/2006/09/034}{\emph{JHEP} {\bf 09} (2006) 034}, [\href{http://arxiv.org/abs/hep-th/0512189}{{\tt hep-th/0512189}}].

\bibitem{Hsu:2006vw}
J.~P. Hsu, A.~Maloney and A.~Tomasiello, \emph{{Black hole attractors and pure spinors}}, \href{http://dx.doi.org/10.1088/1126-6708/2006/09/048}{\emph{JHEP} {\bf 09} (2006) 048}, [\href{http://arxiv.org/abs/hep-th/0602142}{{\tt hep-th/0602142}}].

\bibitem{followup}
{S. Demulder, D. L\"ust and T. Raml}, \emph{{Geometric flows, the Swampland and $N=2$ flux compactifications}}, {\emph{(work in progress)} }.

\bibitem{Delgado:2024skw}
M.~Delgado, D.~van~de Heisteeg, S.~Raman, E.~Torres, C.~Vafa and K.~Xu, \emph{{Finiteness and the Emergence of Dualities}},  \href{http://arxiv.org/abs/2412.03640}{{\tt 2412.03640}}.

\bibitem{Klimcik:1996nq}
C.~Klimcik and P.~Severa, \emph{{NonAbelian momentum winding exchange}}, \href{http://dx.doi.org/10.1016/0370-2693(96)00755-1}{\emph{Phys. Lett. B} {\bf 383} (1996) 281--286}, [\href{http://arxiv.org/abs/hep-th/9605212}{{\tt hep-th/9605212}}].

\bibitem{CaboBizet:2019sku}
N.~Cabo~Bizet, C.~Damian, O.~Loaiza-Brito and D.~M. Pe\~na, \emph{{Leaving the Swampland: Non-geometric fluxes and the Distance Conjecture}}, \href{http://dx.doi.org/10.1007/JHEP09(2019)123}{\emph{JHEP} {\bf 09} (2019) 123}, [\href{http://arxiv.org/abs/1904.11091}{{\tt 1904.11091}}].

\bibitem{Gkountoumis:2023fym}
G.~Gkountoumis, C.~Hull, K.~Stemerdink and S.~Vandoren, \emph{{Freely acting orbifolds of type IIB string theory on T$^{5}$}}, \href{http://dx.doi.org/10.1007/JHEP08(2023)089}{\emph{JHEP} {\bf 08} (2023) 089}, [\href{http://arxiv.org/abs/2302.09112}{{\tt 2302.09112}}].

\bibitem{Rajaguru:2024emw}
M.~Rajaguru, A.~Sengupta and T.~Wrase, \emph{{Fully stabilized Minkowski vacua in the 2$^{6}$ Landau-Ginzburg model}}, \href{http://dx.doi.org/10.1007/JHEP10(2024)095}{\emph{JHEP} {\bf 10} (2024) 095}, [\href{http://arxiv.org/abs/2407.16756}{{\tt 2407.16756}}].

\bibitem{Becker:2024ijy}
K.~Becker, M.~Rajaguru, A.~Sengupta, J.~Walcher and T.~Wrase, \emph{{Stabilizing massless fields with fluxes in Landau-Ginzburg models}}, \href{http://dx.doi.org/10.1007/JHEP08(2024)069}{\emph{JHEP} {\bf 08} (2024) 069}, [\href{http://arxiv.org/abs/2406.03435}{{\tt 2406.03435}}].

\bibitem{Streets:2024rfo}
J.~Streets, C.~Strickland-Constable and F.~Valach, \emph{{Ricci flow on Courant algebroids}},  \href{http://arxiv.org/abs/2402.11069}{{\tt 2402.11069}}.

\bibitem{Andriot:2012an}
D.~Andriot, O.~Hohm, M.~Larfors, D.~L\"ust and P.~Patalong, \emph{{Non-Geometric Fluxes in Supergravity and Double Field Theory}}, \href{http://dx.doi.org/10.1002/prop.201200085}{\emph{Fortsch. Phys.} {\bf 60} (2012) 1150--1186}, [\href{http://arxiv.org/abs/1204.1979}{{\tt 1204.1979}}].

\bibitem{Andriot:2013xca}
D.~Andriot and A.~Betz, \emph{{$\beta$-supergravity: a ten-dimensional theory with non-geometric fluxes, and its geometric framework}}, \href{http://dx.doi.org/10.1007/JHEP12(2013)083}{\emph{JHEP} {\bf 12} (2013) 083}, [\href{http://arxiv.org/abs/1306.4381}{{\tt 1306.4381}}].

\bibitem{Severa:2016lwc}
P.~\v{S}evera and F.~Valach, \emph{{Ricci flow, Courant algebroids, and renormalization of Poisson\textendash{}Lie T-duality}}, \href{http://dx.doi.org/10.1007/s11005-017-0968-5}{\emph{Lett. Math. Phys.} {\bf 107} (2017) 1823--1835}, [\href{http://arxiv.org/abs/1610.09004}{{\tt 1610.09004}}].

\bibitem{mccann2010ricci}
R.~J. McCann and P.~M. Topping, \emph{Ricci flow, entropy and optimal transportation}, {\emph{American Journal of Mathematics} {\bf 132} (2010) 711--730}.

\bibitem{Stout:2021ubb}
J.~Stout, \emph{{Infinite Distance Limits and Information Theory}},  \href{http://arxiv.org/abs/2106.11313}{{\tt 2106.11313}}.

\bibitem{Stout:2022phm}
J.~Stout, \emph{{Infinite Distances and Factorization}},  \href{http://arxiv.org/abs/2208.08444}{{\tt 2208.08444}}.

\bibitem{Witten:2024yod}
E.~Witten, \emph{{Instantons and the Large N=4 Algebra}},  \href{http://arxiv.org/abs/2407.20964}{{\tt 2407.20964}}.

\bibitem{streets2022scalarcurvatureentropygeneralized}
J.~Streets, \emph{Scalar curvature, entropy, and generalized ricci flow},  2022.

\end{thebibliography}\endgroup
\end{document}